\documentclass[twocolumn,twocolappendix]{aastex701}

\usepackage{xcolor}

\newcommand{\rhoHIo}{\rho_{\rm HI,0}}
\newcommand{\MHI}{M_{\rm HI}}
\newcommand{\Mvir}{M_{\rm vir}}
\newcommand{\Mgas}{M_{\rm gas}}
\newcommand{\Rvir}{R_{\rm vir}}
\newcommand{\cNFW}{c_{\rm NFW}}
\newcommand{\rhofive}{\rho_5}

\newcommand{\avgvr}{\langle v_r\rangle}

\begin{document}

\title{Igniting Galaxy Formation in the Postreionization Universe, II: Nature versus Nurture}

\author[orcid=0000-0002-3430-3232,sname='Moreno']{Jorge Moreno}
\affiliation{Department of Physics and Astronomy, Pomona College, Claremont, CA 91711, USA}
\email[show]{jorge.moreno@pomona.edu}

\author[orcid=0009-0000-3376-9869]{Defne Alu\c{c}}
\affiliation{Department of Physics and Astronomy, Pomona College, Claremont, CA 91711, USA}
\email{dais2025@mymail.pomona.edu}

\author[orcid=0009-0008-4545-159X]{Eisha L\'opez}
\affiliation{Department of Physics and Astronomy, Pomona College, Claremont, CA 91711, USA}
\email{edlx2025@mymail.pomona.edu}

\author[orcid=0009-0006-0533-102X]{Vivienne Wood}
\affiliation{Department of Physics and Astronomy, Pomona College, Claremont, CA 91711, USA}
\email{vwvo2025@mymail.pomona.edu}

\author[orcid=0009-0009-9418-7358]{Emmanuel Santiago}
\affiliation{Department of Physics and Astronomy, Pomona College, Claremont, CA 91711, USA}
\email{eszh2022@mymail.pomona.edu}

\author[0000-0002-5908-737X]{Francisco~J. Mercado}
\affiliation{Department of Physics and Astronomy, Pomona College, Claremont, CA 91711, USA}
\affiliation{Department of Physics and Astronomy, University of Southern California, Los Angeles, CA 90089, USA}
\email{francisco.mercado@pomona.edu}

\author[0000-0001-9200-169X]{Caleb~R. Choban}
\affiliation{Department of Astronomy, Indiana University, Bloomington, IN 47405, USA}
\email{cchoban@iu.edu}

\author[0000-0002-2651-7281]{Coral Wheeler}
\affiliation{Department of Physics and Astronomy, California State Polytechnic University, Pomona, Pomona, CA 91768, USA}
\email{cwheeler@cpp.edu}

\author[0000-0003-1848-5571]{M.~Katy Rodriguez Wimberly}
\affiliation{Department of Physics and Astronomy, California State University, San Bernardino, San Bernardino, CA 92407, USA}
\email{maria.wimberly@csusb.edu}

\author[0000-0002-1109-1919]{Robert Feldmann}
\affiliation{Department of Astrophysics, Universit\"at Z\"urich, Zurich, CH-8057, Switzerland}
\email{feldmann@physik.uzh.ch}

\begin{abstract}
Although the ultraviolet background from reionization is thought to conspire against the smallest halos, some gas-rich halos in the post-reionization Universe still form stars.
We ask \emph{which properties, environmental or intrinsic, best discriminate recently-ignited from starless halos}. Using FIREbox,
we compare recently-ignited halos to mass- and gas-matched
\emph{starless} halos at $z=2$--6. We test three families of properties beyond
$\Mvir$, $\Mgas$, and $z$: \emph{geometric} environment (the Perturbation Index and local
galaxy density $\rhofive$), \emph{hydrodynamic} environment (the $1$--$2\,\Rvir$ gas shell), and
\emph{intrinsic} halo properties (concentration and the neutral-gas reservoir). A nested random forest
finds that geometric environment adds little, while the hydrodynamic shell is informative.
Three complementary properties each discriminate ignited from starless halos beyond the
mass--gas--redshift baseline (AUC $=0.633$): total HI mass ($\MHI$, AUC $=0.733$), the neutral-gas radial extent
($r_{\rm HI,50}/\Rvir$, AUC $=0.730$), and the central HI density ($\rhoHIo$, AUC $=0.709$); jointly they
reach AUC $=0.822$, the highest combination tested. At fixed mass and redshift,
recently-ignited halos harbor more HI than matched controls, and their HI radial extent is larger in halos whose first recently formed star particle is older.
The same reservoir tracks recent ignition: from $z=6$ to $z=0$ the median
starless halo keeps its gas ($+0.05$~dex) while its neutral fraction falls by $4.3$~dex, so it
is ionized rather than exhausted and the ignition window closes below $z\simeq2$.
A halo's neutral-gas reservoir discriminates ignited from starless halos
more sharply than its location in the cosmic web --- as discriminators, nature outranks nurture.
\end{abstract}

\keywords{Galaxy formation --- Galaxy environments --- Star formation --- Neutral hydrogen clouds --- Galaxy dark matter halos --- Reionization}

\section{Introduction} \label{sec:intro}
\setcounter{footnote}{0}   %

Whether a galaxy's fate is determined by its intrinsic properties or by its environment --- nature or nurture --- is a long-standing question in the field of galaxy evolution. At the present epoch, halo mass is a prominent driver of star formation and quenching, with large-scale environment and assembly history contributing only weakly at fixed mass \citep{Zu2018}. 
Meanwhile, many authors attribute secondary variations in galaxy and halo properties to \emph{assembly bias}, the dependence of clustering on halo properties beyond mass \citep[e.g. concentration, formation time, and the local tidal field,][]{Gao2005,Wechsler2006,GaoWhite2007,Hahn2009,Feldmann2017,Paranjape2018,Mao2018,Feldmann2019,Xu2021}. Most recently, tidal ``perturbation index'' (PI)
measures have been proposed as environmental drivers of low-mass galaxy structure \citep{Jackson2021,Mercado2025},
building on classical tidal indices \citep{Karachentsev2004} and nearest-neighbor density estimators
\citep{Verley2007,Brough2013,Bradford2015}.

This intrinsic-versus-environment dichotomy takes a sharp form at the \emph{threshold} of galaxy formation. The current view suggests that reionization heats the intergalactic medium and conspires against the smallest halos. Namely, below a redshift-dependent critical
mass, these objects cannot accrete, retain, or cool the gas needed to form stars \citep{Rees1986,Efstathiou92,BarkanaLoeb1999,Sawala2016,Benitez-Llambay2020}. Here, photoheating both
strips gas already bound to the halo and blocks fresh accretion \citep{Gnedin2000,Hoeft2006,Okamoto2008uv,NohMcQuinn2014}.
Such suppression also helps to reconcile the observed abundance of faint satellites with the subhalo mass function from simulations \citep{Bullock2000}.

Framed this way, the ignition threshold is not a single barrier, but a rather complex \emph{pipeline}: gas must first be
accreted, then retained against photoevaporation, then cooled and concentrated, and only then converted into stars \citep[][hereafter, Paper~I]{Moreno2026}. Analytic treatments collapse this chain into one critical mass or circular velocity \citep{Rees1986,Benitez-Llambay2020} \citep[but see][]{Munshi2021}, which captures the net suppression well but cannot establish \emph{which} step is most relevant. By design, Paper~I focuses on the survivors at the end of this pipeline: halos
that had already accreted and retained their gas. This work now takes the next step. Rather than only asking which halos ignite, we also inquire which stage in the chain actually discriminates ignited from starless halos. To accomplish this, we pit \emph{three successive stages} against one another: the large-scale environment a
halo inhabits, the gas supply reaching it through the $1$--$2\,\Rvir$ shell ($\Rvir$ being the halo's virial radius), and the neutral reservoir it has assembled. Each stage feeds the next, so a mechanism acting early should leave a detectable imprint downstream. 

Conventionally, ignition at the galaxy-formation threshold is treated as a phenomenon of the
reionization era itself ($z\gtrsim6$), with halos that fail to ignite frozen thereafter as
reionization relics. In Paper~I we employ \texttt{FIREbox}, a cosmological-volume simulation
\citep{Feldmann2022}, to show instead that ignition remains an
\emph{ongoing} process throughout the postreionization era, continuing down to cosmic noon
($z\simeq2$), so that recently-ignited and still-starless halos coexist across $z=2$--$6$. In that paper, we
find that the properties of the interstellar medium and the structure of the host halo also play a role in
switching on star formation. Paper~I does not, however, address environment. Cosmic-web stripping and tidal interactions can remove or suppress gas from low-mass halos as they thread the filamentary web, shaping
which systems retain the neutral gas that might eventually help ignite new stars \citep{Benitez-Llambay2013,Benitez-Llambay2017,TBL2026}.

Whether large-scale environment traces \emph{recent ignition}, rather than merely rearranging gas, is directly testable. Thus, we ask the following pointed question.
\emph{Among the many properties that distinguish an igniting halo from a matched starless one, which actually discriminate between them: its place in the cosmic web, the gas supply through its own $1$--$2\,\Rvir$ shell, or the intrinsic nature of its own neutral-gas reservoir?} We organize the candidate properties into three
families --- \emph{geometric} environment (the perturbation index and local galaxy density),
\emph{hydrodynamic} environment (the $1$--$2\,\Rvir$ gas-supply shell), and \emph{intrinsic}
properties (concentration and the HI reservoir) --- and rank them using a \emph{matched-control nested regression} (see \S~\ref{subsec:regression} for terminology). Here, we use ``nature'' and ``nurture'' descriptively, for intrinsic and environmental properties, whose ability to discriminate recently ignited from starless halos we compare. Namely, our analysis quantifies associations and does not by itself establish causation (\S~\ref{subsec:regression}).

This manuscript is organized as follows. Section~\ref{sec:methods}
describes the simulation, samples, the three axis families, and an illustrative representative halo pair. Section~\ref{sec:results} presents the three families in turn, their redshift evolution, and the enhancement and
regression synthesis. Sections~\ref{sec:discussion} and~\ref{sec:conclusions} include our discussion and conclusions.

\section{Methods} \label{sec:methods}

\begin{figure*}[!tbp]
\vskip6pt\hbox to\hsize{\hfill\vbox{\parskip=0pt\hsize=0.48\textwidth\includegraphics[width=0.48\textwidth]{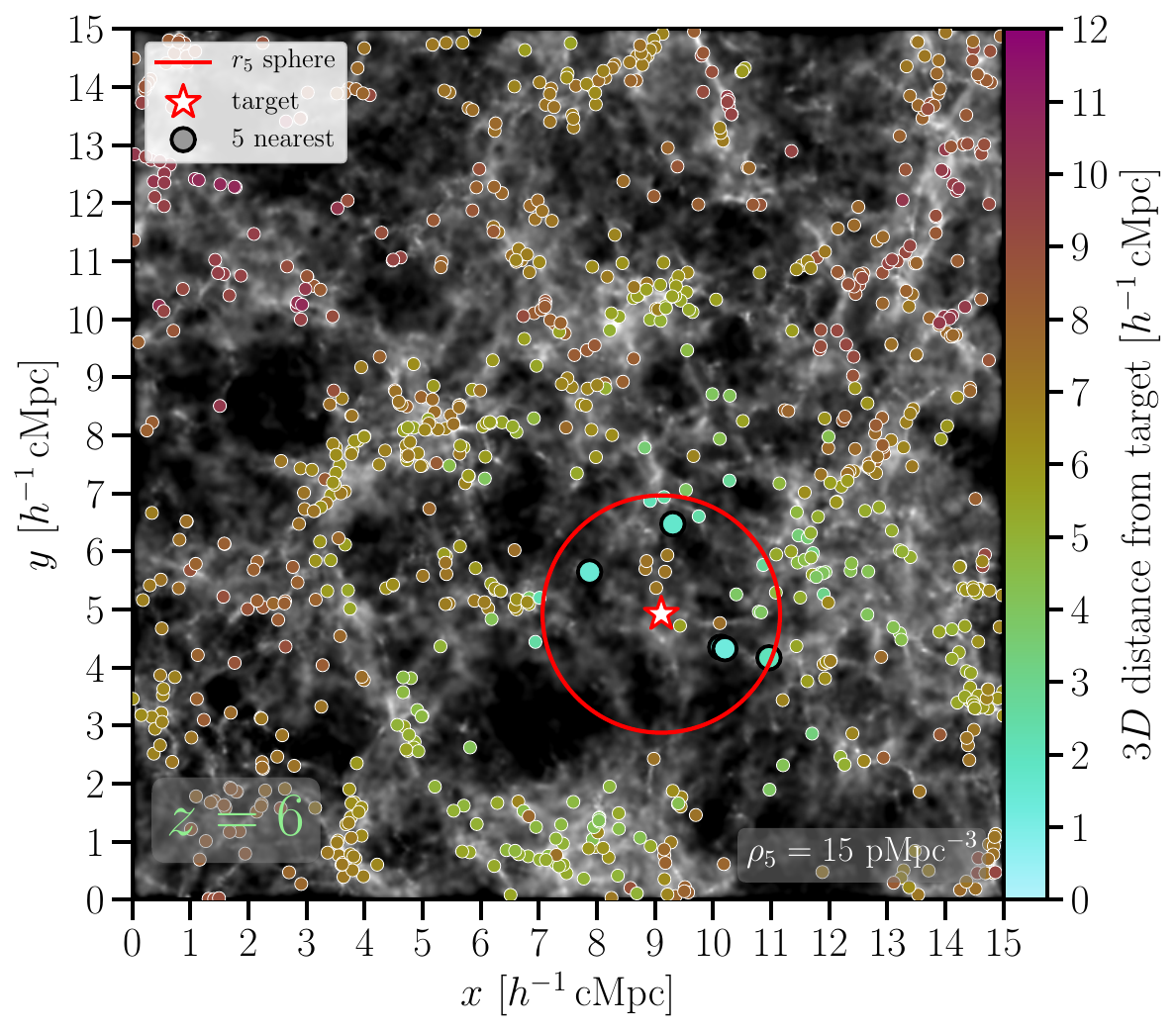}\vskip2pt\vtop{\centering\footnotesize\hsize=0.48\textwidth (a) Ignited, $\rhofive$\vskip1pt}}\hfill
          \hfill\vbox{\parskip=0pt\hsize=0.48\textwidth\includegraphics[width=0.48\textwidth]{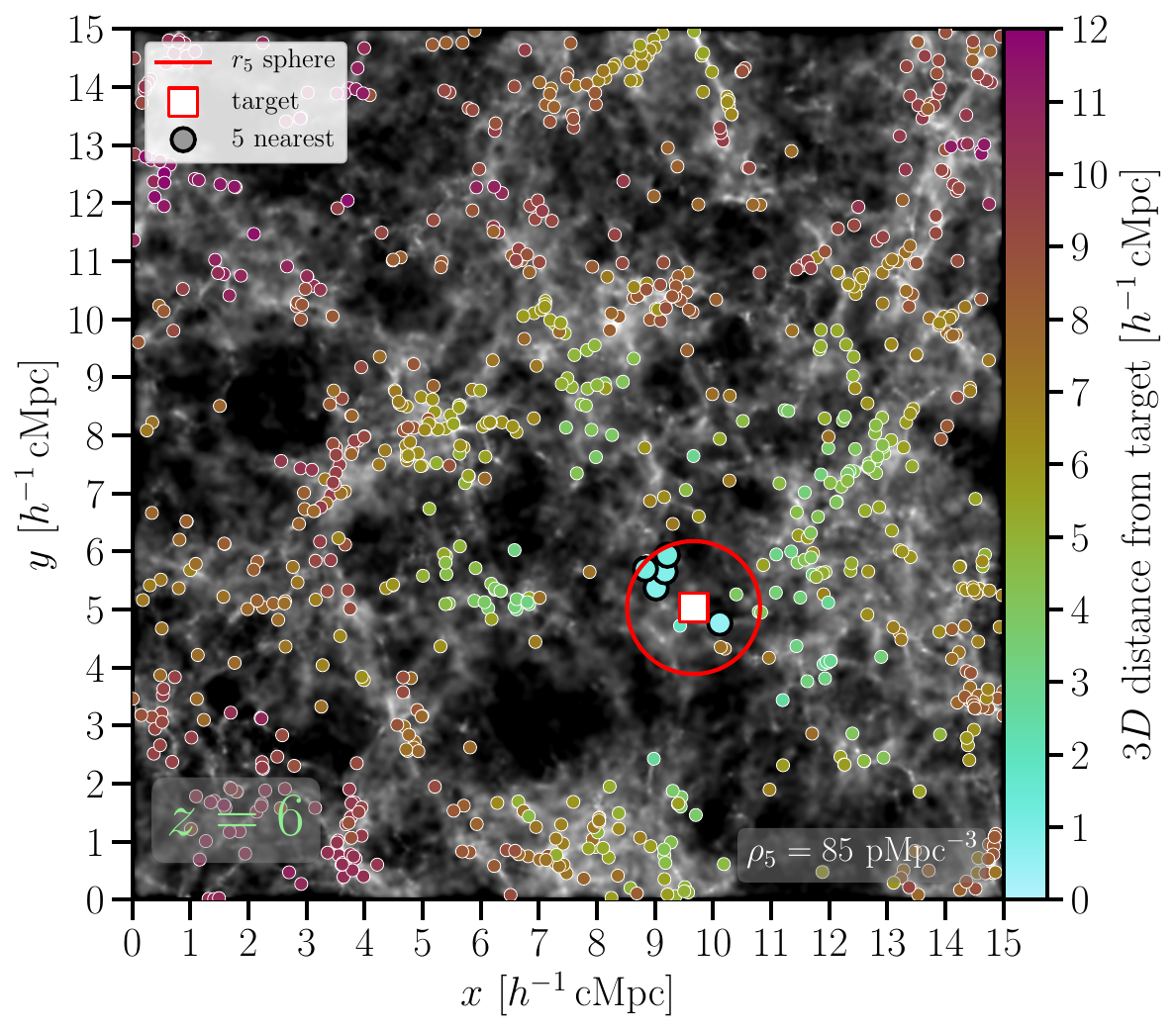}\vskip2pt\vtop{\centering\footnotesize\hsize=0.48\textwidth (b) Starless, $\rhofive$\vskip1pt}}\hfill}\vskip6pt
\vskip6pt\hbox to\hsize{\hfill\vbox{\parskip=0pt\hsize=0.48\textwidth\includegraphics[width=0.48\textwidth]{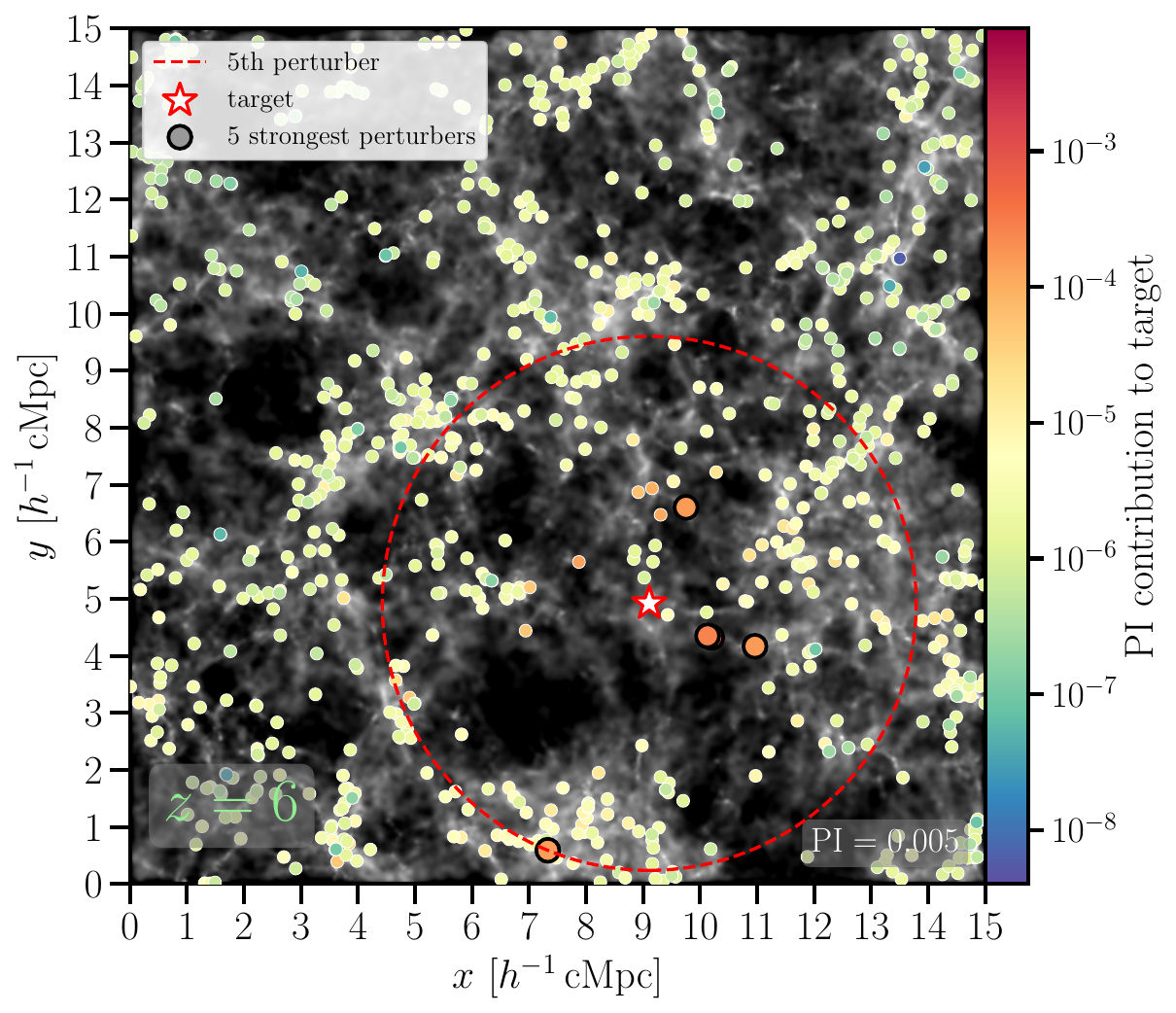}\vskip2pt\vtop{\centering\footnotesize\hsize=0.48\textwidth (c) Ignited, PI\vskip1pt}}\hfill
          \hfill\vbox{\parskip=0pt\hsize=0.48\textwidth\includegraphics[width=0.48\textwidth]{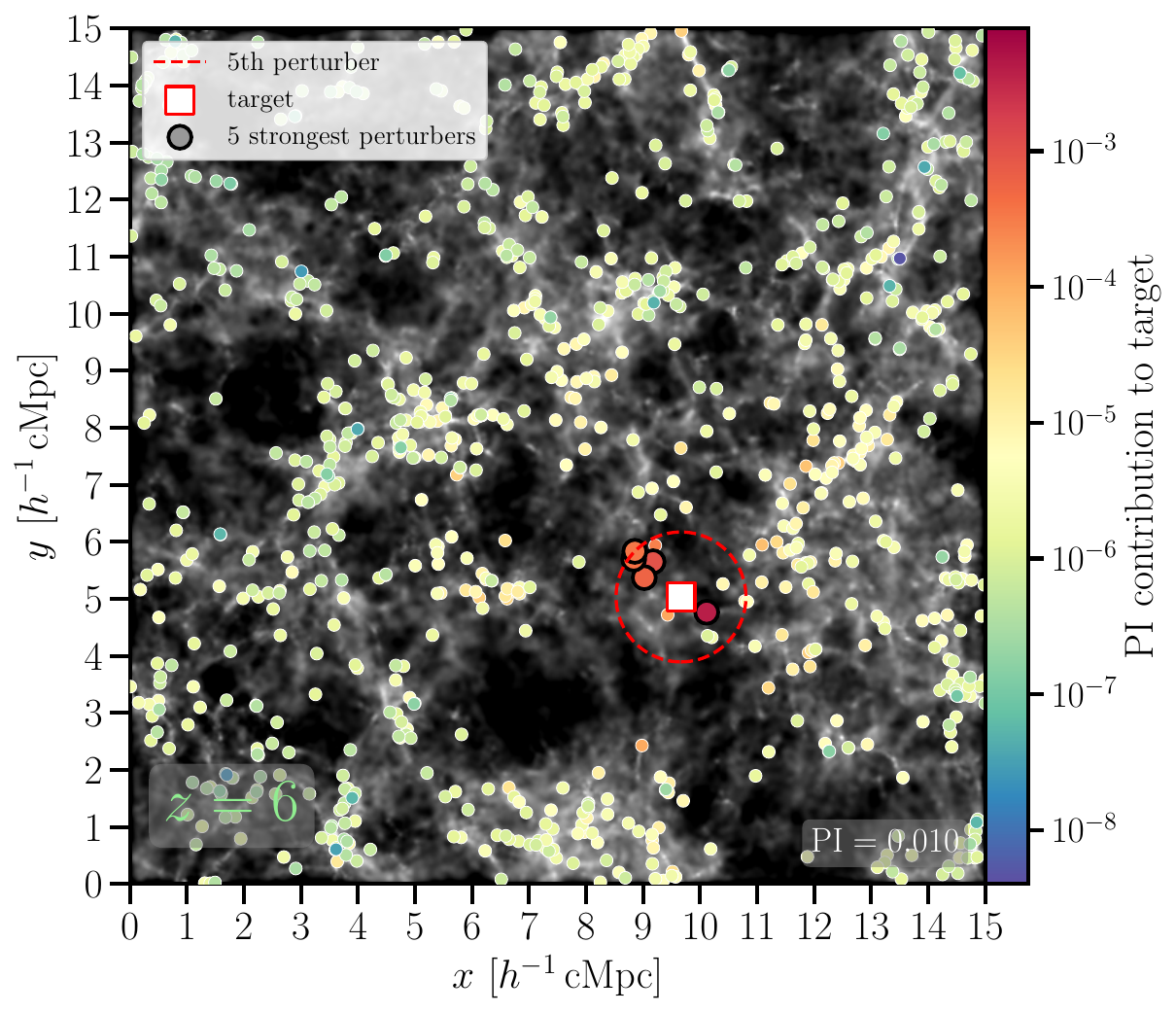}\vskip2pt\vtop{\centering\footnotesize\hsize=0.48\textwidth (d) Starless, PI\vskip1pt}}\hfill}\vskip6pt
\caption{Cosmic-web location of a representative starless-ignited pair at $z=6$, over the gas density field: a
recently-ignited halo (left; star) and its $\Mvir$- and $\Mgas$-matched starless partner (right; square),
matched to $0.009$~dex ($\Mvir=3.9\times10^{8}\,M_\odot$, $\Mgas=2.9\times10^{7}\,M_\odot$). This same
pair appears identically in Figures~\ref{fig:rv} and~\ref{fig:himap}, to describe each family thoroughly
before we turn to the entire population.
\emph{Top:} Local number density ($\rhofive$), with the five nearest galaxies (black edges) and the
$r_5$ solid circle in red. \emph{Bottom:} Perturbation Index (PI), with each companion colored coded by its
contribution; highlighting the five strongest perturbers (black edges), while the dashed red circle marks the
distance to the fifth one.
The ignited halo occupies the \emph{poorer, less perturbed} environment on both measures
($\rhofive = 15$ versus $85\,{\rm pMpc}^{-3}$; ${\rm PI}=5.1\times10^{-3}$ versus
$1.0\times10^{-2}$). The two halos are $1.6$~Mpc apart in three dimensions ($\sim500\,\Rvir$). Both geometric environmental measures are defined in
Section~\ref{subsec:axes} and listed in Table~\ref{tab:quantities}. \label{fig:web}}
\end{figure*}

\subsection{Simulation} \label{subsec:sim}

We employ \texttt{FIREbox} \citep{Feldmann2022}, a cosmological-volume simulation with $22.1\,\mathrm{cMpc}$ on a side,
evolved with the Feedback In Realistic Environments\footnote{\mbox{\url{https://fire.northwestern.edu}}} (\texttt{FIRE-2}) physics model \citep{FIRE2} in the
meshless finite-mass mode of \texttt{GIZMO} \citep{GIZMO}. The simulation resolves the multiphase
interstellar medium with $1024^3$ baryonic and $1024^3$ dark-matter particles, of initial mass $m_{\rm b}=6.3\times10^{4}\,M_\odot$ and $m_{\rm dm}=3.3\times10^{5}\,M_\odot$. Force resolution is fixed at $80$~pc (physical) for dark matter and $12$~pc for star particles. For gas it is set to the adaptive smoothing length, down to a minimum of $1.5$~pc, reached only in the densest regions. The simulation follows radiative heating and cooling under a spatially uniform \citet{CAFG2009} ultraviolet/X-ray background with local
self-shielding, restricts star formation to self-gravitating, self-shielded, Sobolev/Jeans-unstable molecular
gas, and injects stellar feedback from supernovae, stellar mass loss, radiation pressure, and
photoheating. We identify central halos with the \texttt{AMIGA Halo Finder} \citep{Knollmann2009},
adopt the \citet{Bryan1998} virial definition, and access the particle data through \texttt{yt}
\citep{Turk2011}. We assume the simulation cosmology ($\Omega_{\rm m}=0.3089$, $\Omega_\Lambda=0.6911$,
$\Omega_{\rm b}=0.0486$, $h=0.6774$) and refer the reader to Paper~I, \citet{FIRE2}, and
\citet{Feldmann2022} for details.

\subsection{Halo samples} \label{subsec:samples}

We adopt the same halo samples as Paper~I and summarize them here. We restrict our analysis to central halos
resolved by at least $128$ dark-matter particles and possessing a resolved gaseous component
($N_{\rm gas}\geq128$). A halo is \emph{starless} when it contains no star particle ($N_\star=0$), and
\emph{recently ignited} when it contains exactly one star particle in total ($N_\star=1$ over the whole halo), that particle lies within $0.15\,\Rvir$ of the halo center, and its
age is below $100\,\mathrm{Myr}$, so that we capture the halo close to the onset of star formation. This yields $192$ recently-ignited halos (hereafter simply
\emph{ignited}) and a parent population of $18{,}284$ starless halos, both across $z=2$--$6$ (in this paper series we examine $z=0,0.5,1,2,3,4,5, 6$).

For each ignited halo we construct a \emph{control} sample of starless halos at the same redshift, matched
in virial and gas mass to within $0.1\,\mathrm{dex}$. Control quantities are averages over this set, and the \emph{enhancement} of a property $x$ is $x/\langle x^{\rm control}\rangle$; values below unity denote \emph{suppression}. Similarly, the \emph{residual} of a property $\log y$, $\Delta\log y$, is its offset from the relation followed by the starless population against $x$, evaluated at fixed $x$.

\subsection{Three families of ignition discriminants} \label{subsec:axes}

\begin{figure}[!tbp]
\includegraphics[width=\linewidth]{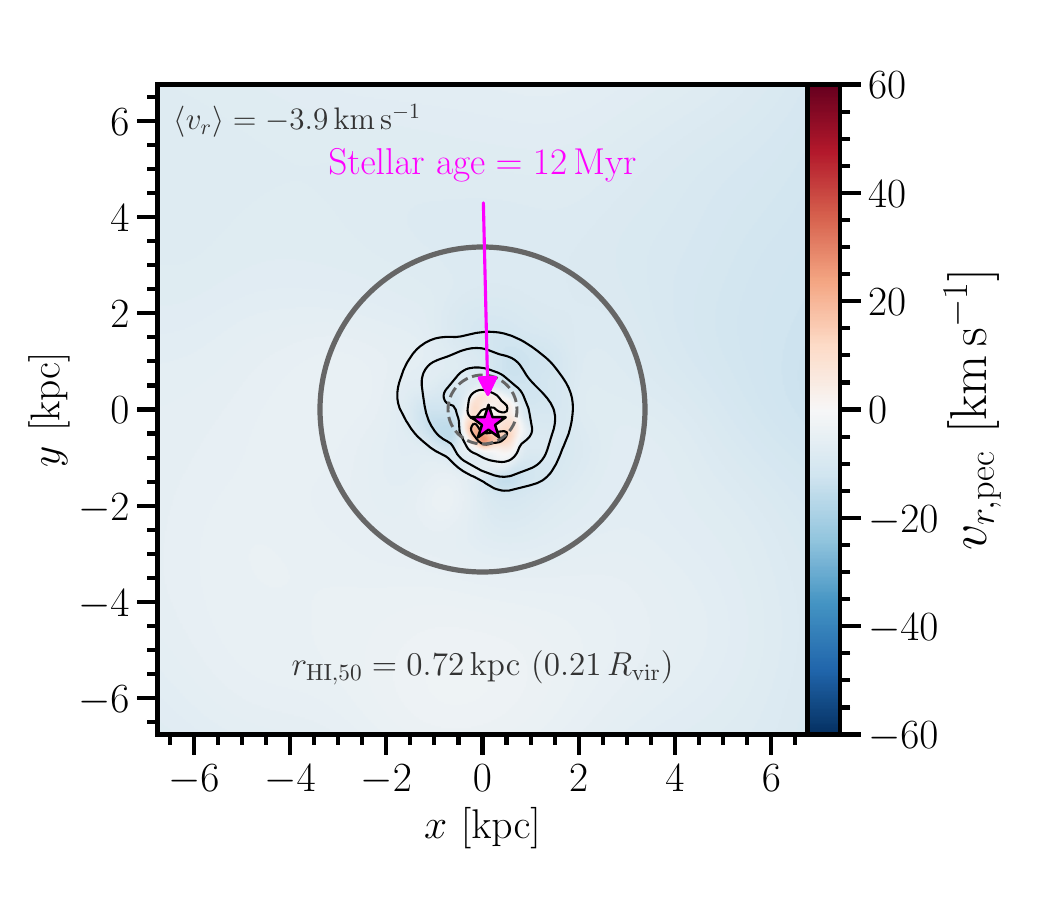}\\[-9pt]
\includegraphics[width=\linewidth]{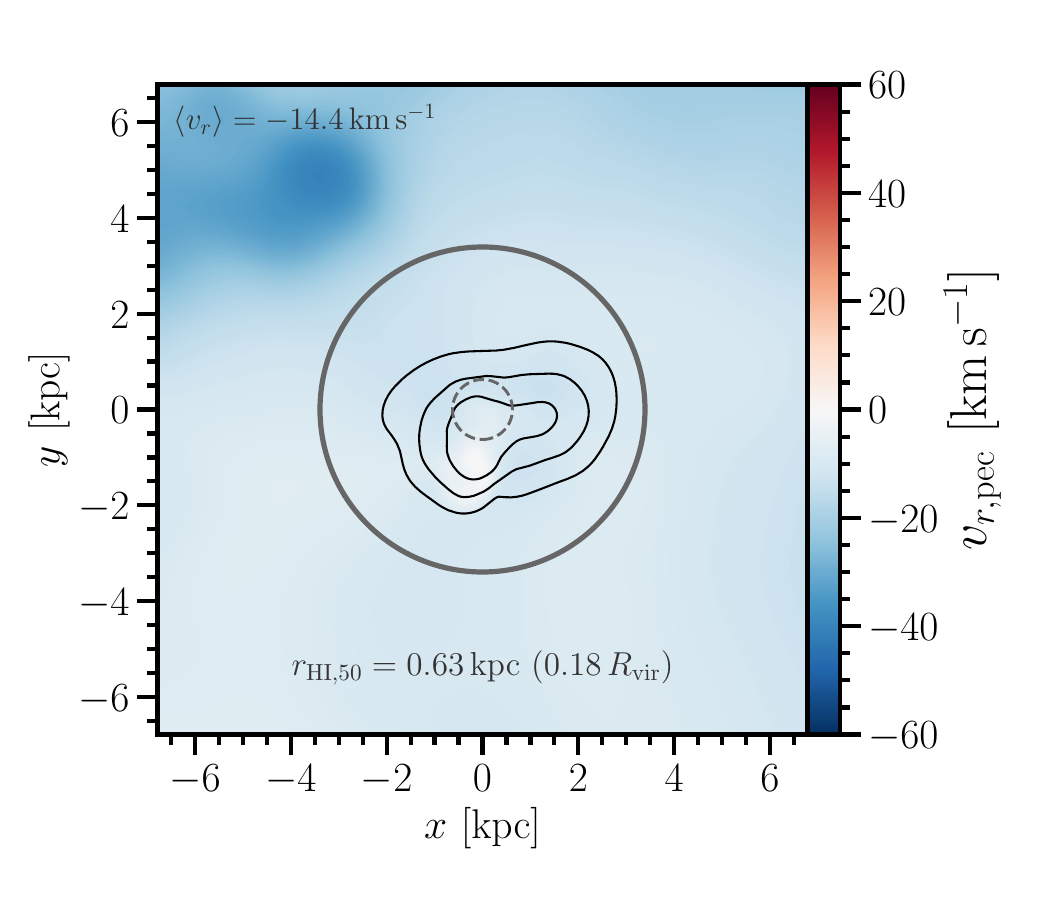}
\caption{Peculiar radial-velocity maps of our representative pair at $z=6$ (blue inflow, red outflow),
within $2\Rvir$: the recently-ignited halo (\emph{top}) and its matched starless partner
(\emph{bottom}). Gray circles: $\Rvir$ (solid) and the HI half-mass radius
$r_{\rm HI,50}$ (dashed). Black contours are HI column density (see Figure~\ref{fig:himap}). The magenta star marks the newly formed stellar
particle ($12$~Myr old), absent in the starless halo. Each panel quotes $\avgvr$, the mass-weighted mean radial velocity over the $1$--$2\,\Rvir$ shell
($-14.4$ versus $-3.9\,{\rm km\,s^{-1}}$, starless against ignited). Both systems are embedded in an inflowing envelope,
so neither is short of a gas supply --- but the \emph{starless} halo has the stronger inflow
($\dot M_{\rm in}=0.26$ versus $0.23\,M_\odot\,{\rm yr^{-1}}$) and the larger shell reservoir ($M_{\rm shell}=3.0\times10^{7}$ versus $2.2\times10^{7}\,M_\odot$). Section~\ref{subsec:axes} and
Table~\ref{tab:quantities} provide these shell conventions. \label{fig:rv}}
\end{figure}

\begin{figure}[!tbp]
\includegraphics[width=\linewidth]{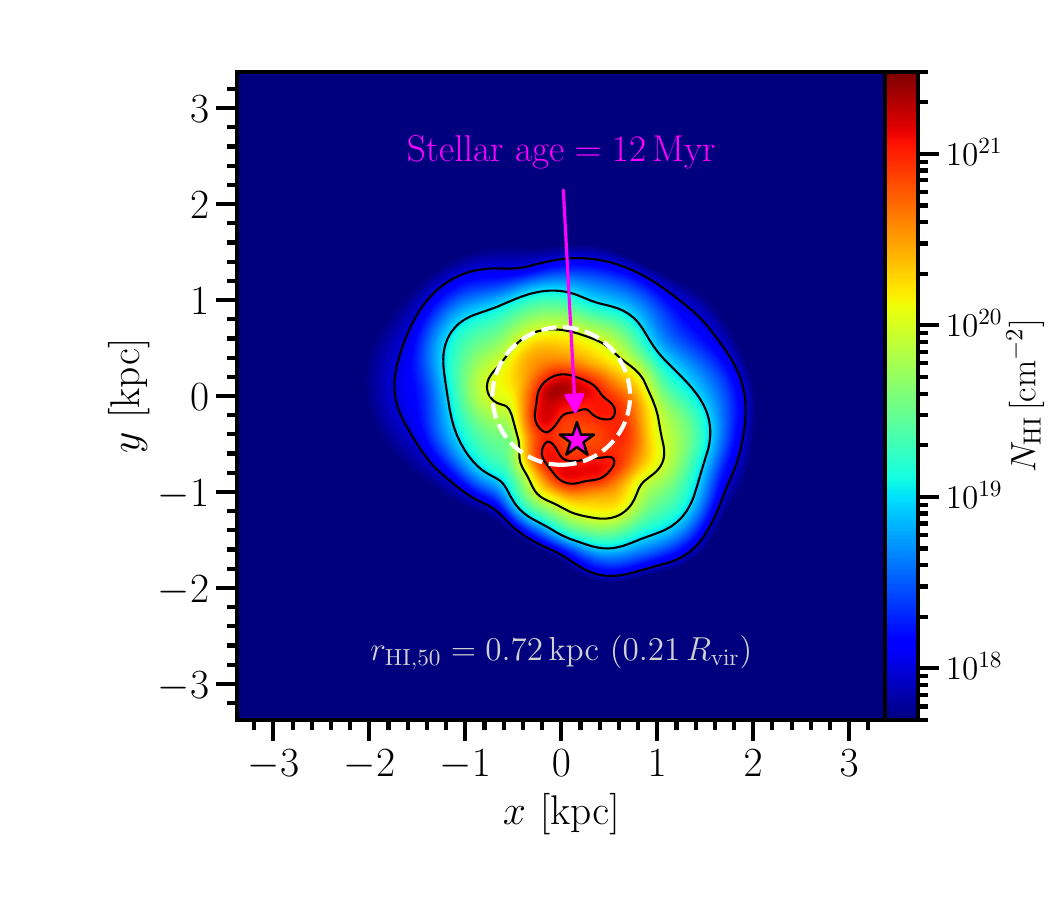}\\[-9pt]
\includegraphics[width=\linewidth]{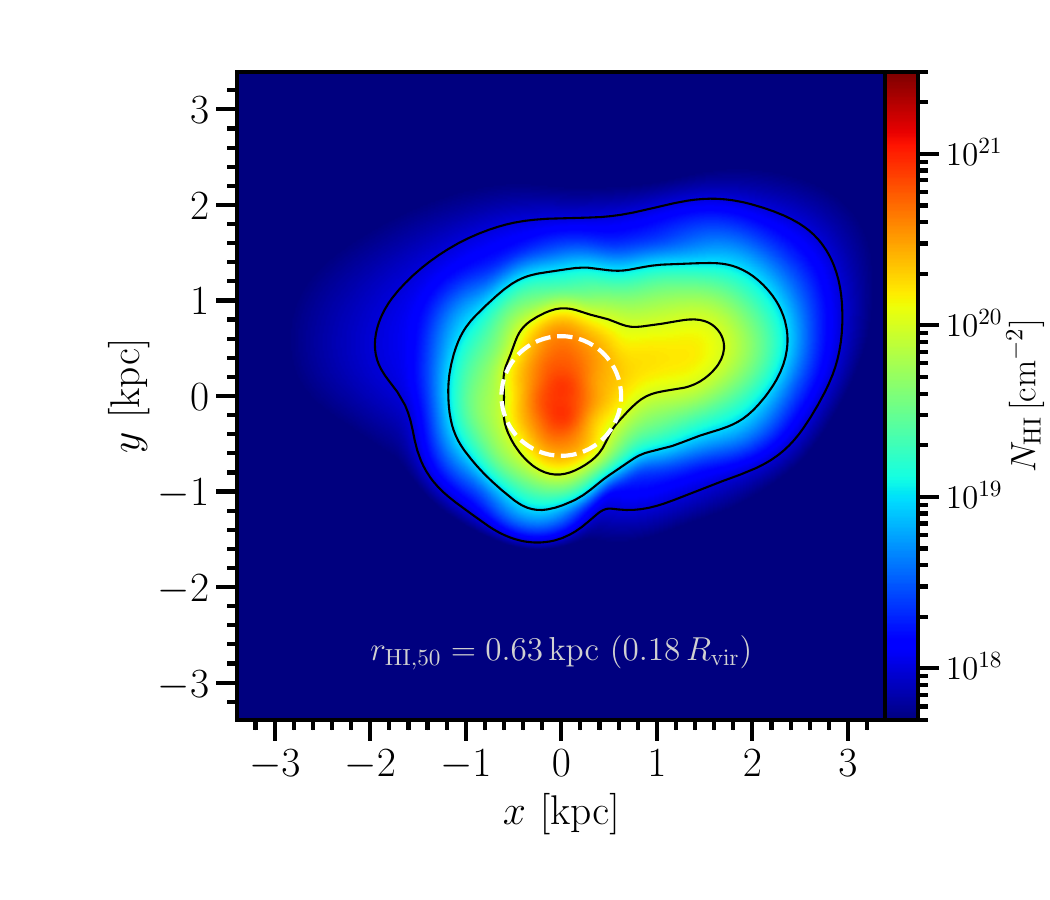}
\caption{HI column-density maps of our representative pair at $z=6$, within $\Rvir$: the recently-ignited
halo (\emph{top}) and its matched starless partner (\emph{bottom}). Both are centered on the
\emph{HI density peak} located by a shrinking-sphere centroid method. 
The ignited halo holds
$1.8\times$ more neutral gas ($\MHI=1.3\times10^{7}$ versus $7.0\times10^{6}\,M_\odot$) and reaches a
higher central density ($\rhoHIo=7.2\times10^{6}$ versus $4.3\times10^{6}\,M_\odot\,{\rm kpc^{-3}}$),
and its newly formed stellar particle is marked. The dashed white circle is the HI half-mass radius
$r_{\rm HI,50}$ about that same peak, larger in the ignited halo ($0.72$~kpc, $0.21\,\Rvir$) than in its
starless partner ($0.63$~kpc, $0.18\,\Rvir$). Black contours mark fixed levels of HI column density $N_{\rm HI}$; the ignited halo's core reaches
$N_{\rm HI}\simeq2\times10^{21}\,{\rm cm^{-2}}$. These reservoir quantities are
defined in Section~\ref{subsec:axes} and collected in Table~\ref{tab:quantities}. \label{fig:himap}}
\end{figure}

We test three families as candidate discriminants of ignition beyond virial mass, gas mass, and redshift. We illustrate each using a \emph{representative} pair of halos in Figures~\ref{fig:web}--\ref{fig:himap} (we analyze the entire population in \S~\ref{sec:results}). Properties within each family are defined below (and included in Table~\ref{tab:quantities}, Appendix~\ref{app:quantities}, for reference).

\begin{itemize}
\setlength{\itemsep}{0.10cm}
\setlength{\parsep}{0pt plus 0pt minus 0pt}
\setlength{\topsep}{0pt plus 0pt minus 0pt}
\item \textbf{Geometric environment} measures where a halo is located among its neighbors. The perturbation index
$\mathrm{PI}=\sum_{c}(M_{\rm vir}^{c}/M_{\rm vir}^{p})(2R_{\rm vir}^{p}/D_{cp})^{3}$
\citep{Verley2007,Mercado2025} sums the tidal influence of surrounding galaxies $c$ on the halo $p$ at
three-dimensional separation $D_{cp}$, while the local density $\rhofive=5/V_5$ is the galaxy number
density within the sphere of radius $r_5$ that reaches the fifth-nearest neighbor, of volume $V_5=(4\pi/3)r_5^3$ \citep{Verley2007}, illustrated for our representative pair in Figure~\ref{fig:web}. \item \textbf{Hydrodynamic environment} measures the gas supply through the surrounding $1$--$2\,\Rvir$ shell: the shell-gas mass ($M_{\rm shell}$), the mean radial velocity ($\avgvr$; mass-weighted,
bulk-subtracted and Hubble-flow corrected, so that $\avgvr<0$ denotes net inflow), and the inflow rate ($\dot M_{\rm in}$), illustrated in Figure~\ref{fig:rv}. \item \textbf{Intrinsic properties} describe the halo and its own gas: the concentration $\cNFW$ \citep{NFW}, taken from \textsc{ahf} in the \citet{Prada2012} form; the total
neutral-hydrogen mass $\MHI$; the
HI extent $r_{\rm HI,50}/\Rvir$ (the neutral-gas half-mass radius modulo virial radius, in units chosen to remove redshift evolution; Appendix~\ref{app:quantities}); and the central HI density $\rhoHIo$, illustrated in Figure~\ref{fig:himap}.
\end{itemize}

\subsection{Nested-model regression (machine learning)} \label{subsec:regression}

To rank the members of these three families we train a \emph{random-forest classifier} \citep{Breiman2001}, as implemented in \texttt{scikit-learn} \citep{Pedregosa2011}, to separate ignited from starless halos. For fuller introductions to these methods, \citet{Acquaviva2023}, \citet{Baron2019}, \citet{HuertasCompany2023} and \citet{Ting2025} are written for astronomers, while \citet{Mehta2019} treats ensemble methods such as random forests in particular detail. 

Our comparison employs \emph{nested} feature sets. Namely, beginning from a \emph{baseline} of virial mass, gas mass and
redshift (as in Paper I), we add each candidate property (or combination thereof) in turn and record the change in
classification performance. Random forests have been used this way before in astronomy to rank the properties associated with
quenching in observed galaxies by weighing local, global and environmental properties against one
another \citep{Bluck2020,Piotrowska2020,Bluck2022}. We apply the same logic here to inquire whether a (gas-resolved) halo
forms stars at all. This comparison does not require the three families to be independent -- each candidate is scored against the same shared baseline, not against one another, so a nested gain stays interpretable even where two families are correlated. Indeed, Section~\ref{subsec:reframe} shows that environment does measurably shape the gas supply, though too weakly to propagate into a comparable signal for ignition itself; Section~\ref{subsec:rotation} confirms the family-level enhancements are in turn close to independent (rank correlations $\lesssim0.07$).

Performance is quantified using the \emph{area under the receiver-operating-characteristic
curve} (AUC), averaged over $100$ random partitions of the sample into $30\%$ for \emph{training} and $70\%$
for \emph{testing}. Those partitions are \emph{class-stratified}. Namely, each preserves the ratio of ignited to
starless halos, so no split is left with too few ignited objects to score. We deliberately adopt this large test fraction because although a more conventional split would give the forest more data to
train on -- with only $192$ ignited halos, a correspondingly smaller test set would give an unstably
estimated AUC. We prefer a stable AUC score over a marginally better-trained forest.

For readers unfamiliar with this measure, the AUC has a simple interpretation. It represents the probability
that a randomly chosen ignited halo is assigned a higher score than a randomly chosen starless one. A
model with no information scores $0.5$, while a perfect classifier achieves $1.0$. Similarly, a score of $0.75$
means that the randomly-chosen ignited halo is ranked higher three times out of four. 

We note that only \emph{differences} in AUC matter here, not absolute values. Namely, we are asking how much a candidate property adds to
what mass, total gas content and redshift (our baseline) already provide. The quantity of interest is therefore
 $\Delta$AUC, the \emph{increment} above this baseline (which is $0.633$ in this work). We also quantify the \emph{uncertainty} on this
increment in two ways. Drawing a fresh $30\%/70\%$ train/test split of the same halos sets a split-to-split \emph{floor} below which two models
should be regarded as indistinguishable (\S~\ref{subsec:geometric}). However, this floor does not
capture the uncertainty of having only $192$ ignited halos available. To capture this, we instead bootstrap our test set, scoring every model on the same resample so that the comparison is apples to
apples. This properly quantifies that uncertainty as a $95\%$ interval on a difference in AUC
(\S~\ref{subsec:synthesis}).

Also, $\rhoHIo$ and $M_{\rm shell}$ will enter our regression scheme as a \emph{residual} rather than just a raw value, because both scale with the halo's own gas content. For $\rhoHIo$ this is $\Delta \log\rhoHIo$, the offset of $\log\rhoHIo$ from the running median of the starless halos at the same $\MHI$. Similarly, $\Delta \log M_{\rm shell}$ is the offset of $\log M_{\rm shell}$ from the $M_{\rm shell}$--$\Mgas$ relation. Expressing a property as a residual about
a population's scaling relation, rather than as a raw value \citep[][]{Mercado2025}, allows us to ask whether
a halo is unusual \emph{for its own gas content} rather than unusual outright. We use a running median
for $\rhoHIo$ because the starless $\MHI$--$\rhoHIo$ relation saturates and is poorly fit by a single power law. For $M_{\rm shell}$, a single
power law fits well across the full range (and it agrees well with its running-median version) at $\log M_{\rm shell}=0.45\log\Mgas+4.13$ ($0.12$~dex scatter).

Lastly, we note that our nested regression measures association, not causation. Namely, a larger $\Delta$AUC means that a property better separates recently ignited from starless halos at fixed $\Mvir$, $\Mgas$ and $z$. However, it does not mean that the property causes ignition. Indeed, two considerations temper any causal reading. First, the HI quantities ($\MHI$, $r_{\rm HI,50}/\Rvir$, $\rhoHIo$) are measured at the snapshot at which a halo is identified as recently ignited, hence close in time to the event, whereas $\rhofive$ and PI characterize the larger-scale surroundings and the shell the gas outside $\Rvir$. A quantity measured close to the onset can discriminate well without being the more fundamental cause. Second, a matched-control enhancement is an association at fixed mass, gas mass and redshift. Meanwhile, correlated properties that we do not control in this analysis, such as assembly history, may contribute. We therefore use ``nature'' and ``nurture'' descriptively, for intrinsic and environmental properties, and we phrase our conclusions in terms of what distinguishes recently ignited halos from starless ones, not what determines ignition.

\section{Results} \label{sec:results} With the three families and our regression machinery in place, we now turn to the full starless/ignited populations. Our goal is to investigate
which stage of the pipeline actually discriminates ignited halos from starless ones: large-scale environment, external/neighboring gas
supply, or the intrinsic HI reservoir a halo has already assembled. The next four subsections (\S\S~\ref{subsec:rotation}--\ref{subsec:intrinsic}) characterize each family in turn using simple matched-control enhancement statistics; the AUC-based ranking of Section~\ref{subsec:regression} is not applied until Section~\ref{subsec:singles}, once all three families have been introduced individually.

\begin{figure}[!tbp]
\includegraphics[width=\linewidth]{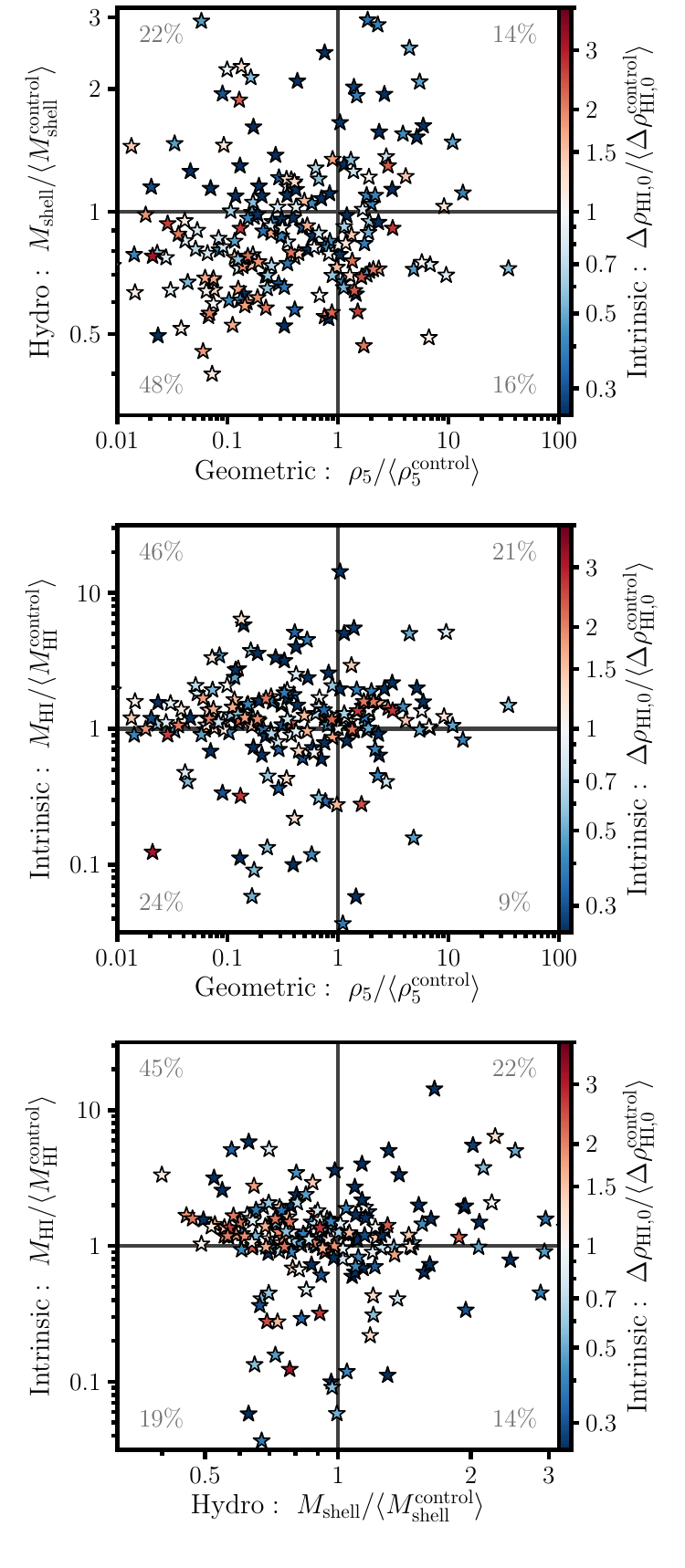}
\caption{The three families in enhancement space, relative to $\Mvir$- and $\Mgas$-matched starless controls
($N=191$; one ignited halo has zero bound HI within its own aperture and is omitted). Axes: geometric ($\rhofive$) and hydrodynamic ($M_{\rm shell}$) environment, plus intrinsic content ($\MHI$);
each as the ratio $x/\langle x^{\rm control}\rangle$ on a logarithmic scale.
Color represents the intrinsic structural residual
$\Delta\rhoHIo/\langle\Delta\rhoHIo^{\rm control}\rangle$, the displacement from the starless
$\MHI$--$\rhoHIo$ sequence.
Solid lines mark unity; each corner reports the share of ignited halos falling in that quadrant, the four summing to $100\%$. Table~\ref{tab:quantities}
lists the quantities behind each axis. \label{fig:rotation}}
\end{figure}

\begin{figure}[!tbp]
\includegraphics[width=\linewidth]{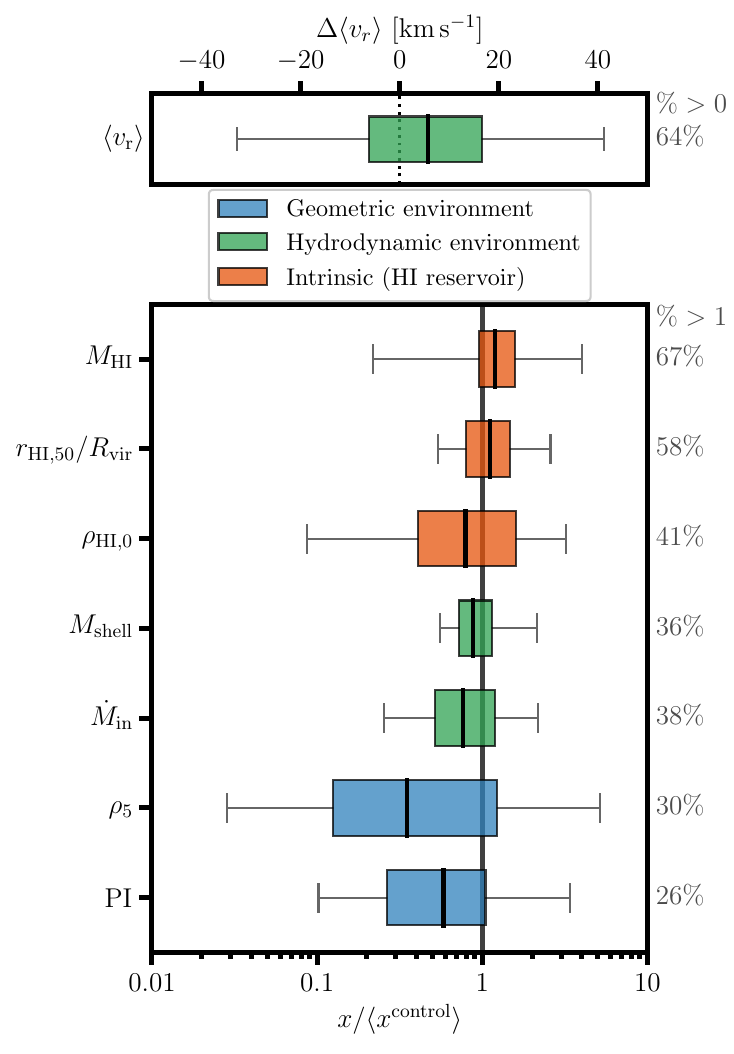}
\caption{Enhancement of recently-ignited halos relative to their $\Mvir$- and $\Mgas$-matched starless controls
($N=192$ for every row except the central HI density, $N=191$: one halo has zero bound HI within its own $0.15\,\Rvir$), grouped by family: geometric (blue), hydrodynamic (green) and intrinsic
(orange). Boxes span the interquartile range with the median marked; whiskers reach the $5$th--$95$th
percentiles. The right-hand column displays the percentage of ignited halos enhanced over their controls: above the
unity line for the seven ratios ($\%>1$), and above zero for $\avgvr$ ($\%>0$). Because $\avgvr$
changes sign, it is presented as a signed difference in ${\rm km\,s^{-1}}$ rather than a ratio, so it occupies the
upper panel, on its own scale. %
The local density ($\rhofive$) and the perturbation index (PI) are the
most suppressed quantities, with the total HI mass ($\MHI$) being the most enhanced. Full distributions in Figure~\ref{fig:envnull} (Appendix~\ref{sec:appendix}). \label{fig:enhflow}}
\end{figure}

\subsection{The three families across the entire population} \label{subsec:rotation}
Before presenting our nested-regression analysis (\S~\ref{subsec:regression}), we compare ignited halos versus carefully-matched controls, following Paper I. Concretely, Figure~\ref{fig:rotation} compares matched-control enhancements for one representative quantity per family. We color-code each panel by the enhancement in $\Delta\log\rhoHIo$, the residual displacement from the starless $\rhoHIo$--$\MHI$ sequence.

Of the three panels, the $\rhofive$--$\MHI$ plane pits geometric environment directly against the
intrinsic HI reservoir, comparing how well intrinsic (nature) and environmental (nurture) properties discriminate. We find that $46\%$ of ignited halos fall in the environment-suppressed, HI-rich quadrant. Turning to the panels that include $M_{\rm shell}$, the color varies systematically only along that hydrodynamic environment axis
(rank correlation $-0.38$), not along the geometric one. Namely, halos with more shell gas than what their own internal gas-mass content would imply sit further
\emph{below} the starless $\MHI$--$\rhoHIo$ sequence. In other words, ignited halos harbor \emph{more} neutral gas than their controls ($67\%$ above), while that gas is
\emph{less} centrally concentrated relative to what that relation predicts (median $-0.40$~dex).

We also note that the enhancements are only weakly correlated (the rank correlation between the geometric and
hydrodynamic axes is $+0.21$, and between geometric and intrinsic $+0.01$). This means that the families are not proxies for one
another. However, both environmental measures point in the same direction. Namely, $70\%$ of ignited halos are suppressed in
$\rhofive$ and $64\%$ in $M_{\rm shell}$ relative to their controls.

Figure~\ref{fig:enhflow} compares net enhancement/suppression values for individual properties -- grouped (and color-coded) by family (geometric and hydrodynamic environment, plus intrinsic). It extends this pattern to every individual property, not just the three representative quantities of Figure~\ref{fig:rotation}. We find that the geometric quantities sit below the unity
line, and the shell mass and inflow rate sit close to it. The mean radial velocity appears to be an
exception, enhanced in $64\%$ of pairs, but a higher, less negative $\avgvr$ means \emph{weaker}
infall. This is in agreement with the other two hydrodynamic properties, suggesting that ignited halos are the less well fed. 

Meanwhile, the total
HI mass stands out, enhanced over its controls in $67\%$ of pairs, followed by the ($\Rvir$-scaled) half-mass radius
($58\%$). The central HI density's \emph{raw} value, by contrast, is essentially unenhanced ($41\%$).
As in Figure~\ref{fig:rotation}, it is the density's deviation from the starless relation (the residual), not its raw value, that discriminates ignited from starless halos. Recently-ignited halos are therefore not distinguished by richer environments (if anything, they inhabit poorer ones), but still contain a heftier neutral-gas~reservoir.

\subsection{Geometric environment} \label{subsec:geometric}
At fixed virial mass, gas mass and redshift, we find that only $30\%$ of ignited halos inhabit a denser
environment than their matched controls, and only $26\%$ exist in a more perturbed one
(Figure~\ref{fig:enhflow}), with median offsets of $-0.46$~dex in $\rhofive$ and $-0.24$~dex in PI respectively. In other words, ignited halos are marginally more isolated. However, this does not necessarily imply that isolation promotes
ignition. Rather, a more robust statement is that \emph{geometric environment, as captured by $\rhofive$ and PI, carries little information about which halo ignites}, far less than the halo's own neutral reservoir would deliver (more below).

\subsection{Hydrodynamic environment} \label{subsec:hydro}
Across the matched sample, ignited halos contain marginally \emph{less} shell gas and weaker inflow
than their starless counterparts. Namely, $36\%$ exceed
their controls in $M_{\rm shell}$ and $38\%$ in $\dot M_{\rm in}$ (Figure~\ref{fig:enhflow}). Likewise, the mean radial velocity is higher (less strongly infalling) in $64\%$ of ignited halos, by a median of
$+5.7$$\,{\rm km\,s^{-1}}$, suggesting that recently-ignited halos are not the ones being fed more vigorously by their surroundings.

\subsection{The intrinsic HI reservoir} \label{subsec:intrinsic}
Ignited halos exceed their controls in $\MHI$ for $67\%$ of matched pairs, and in $r_{\rm HI,50}/\Rvir$ for
$58\%$ (Figure~\ref{fig:enhflow}). In other words, they hold more neutral gas than their starless counterparts, and
spread it over a larger radius. Unlike the other two families, whose apparent discriminating power was
either negligible or borrowed from a correlation with this same reservoir, both effects here are
real, which is why this family merits closer inspection.

\begin{figure*}[!tbp]
\includegraphics[width=0.5\textwidth]{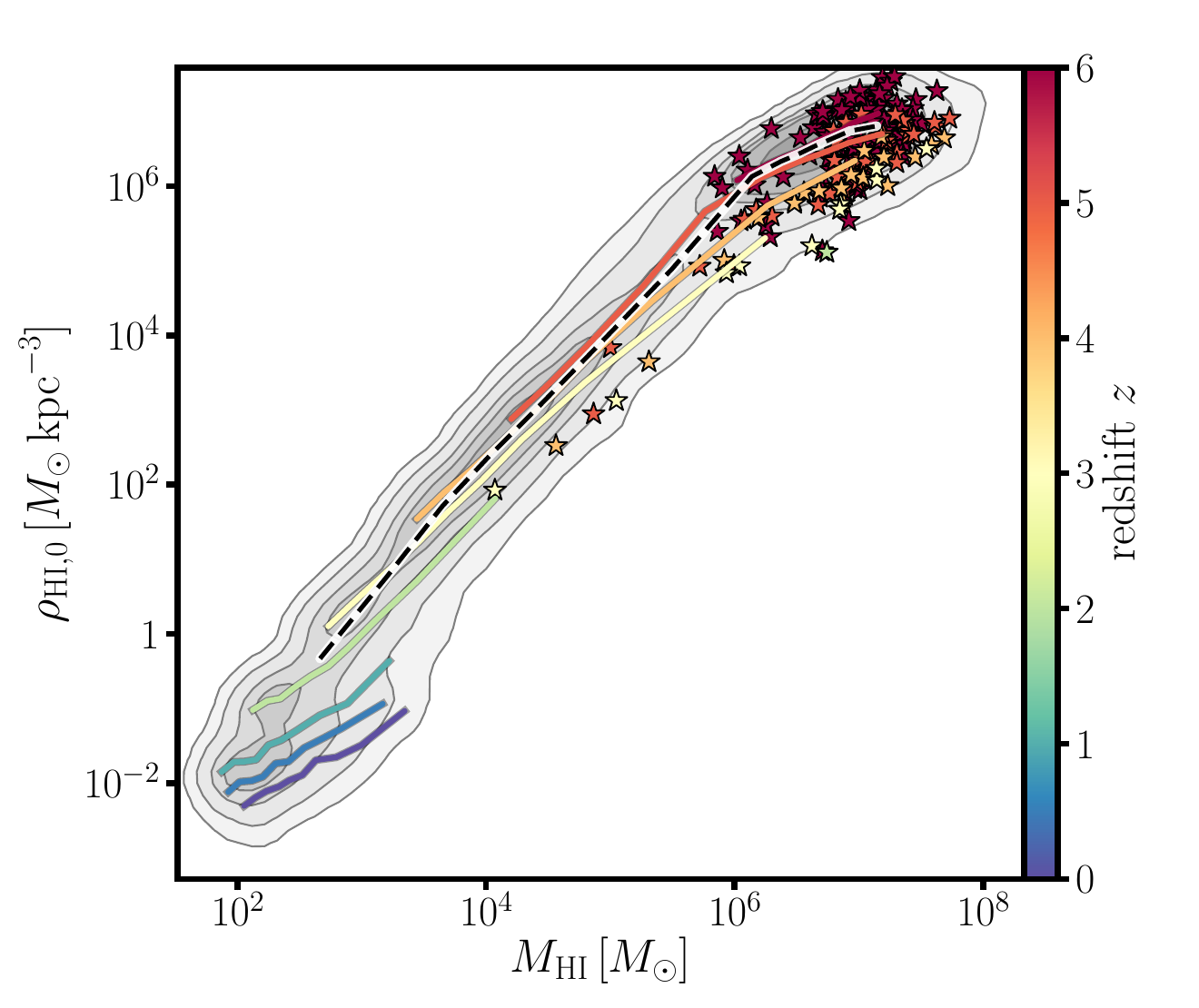}%
\includegraphics[width=0.5\textwidth]{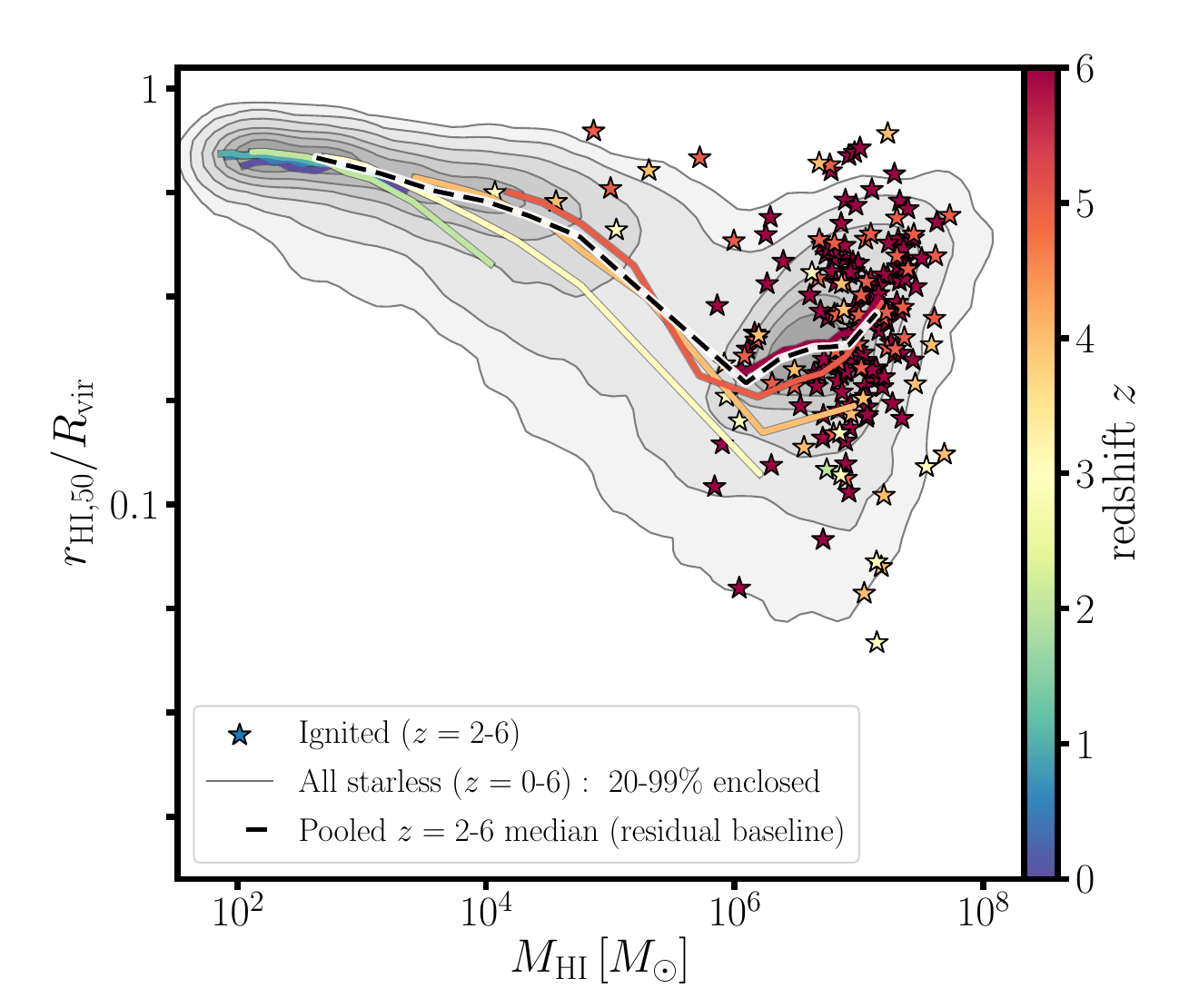}\\[2pt]
\includegraphics[width=0.5\textwidth]{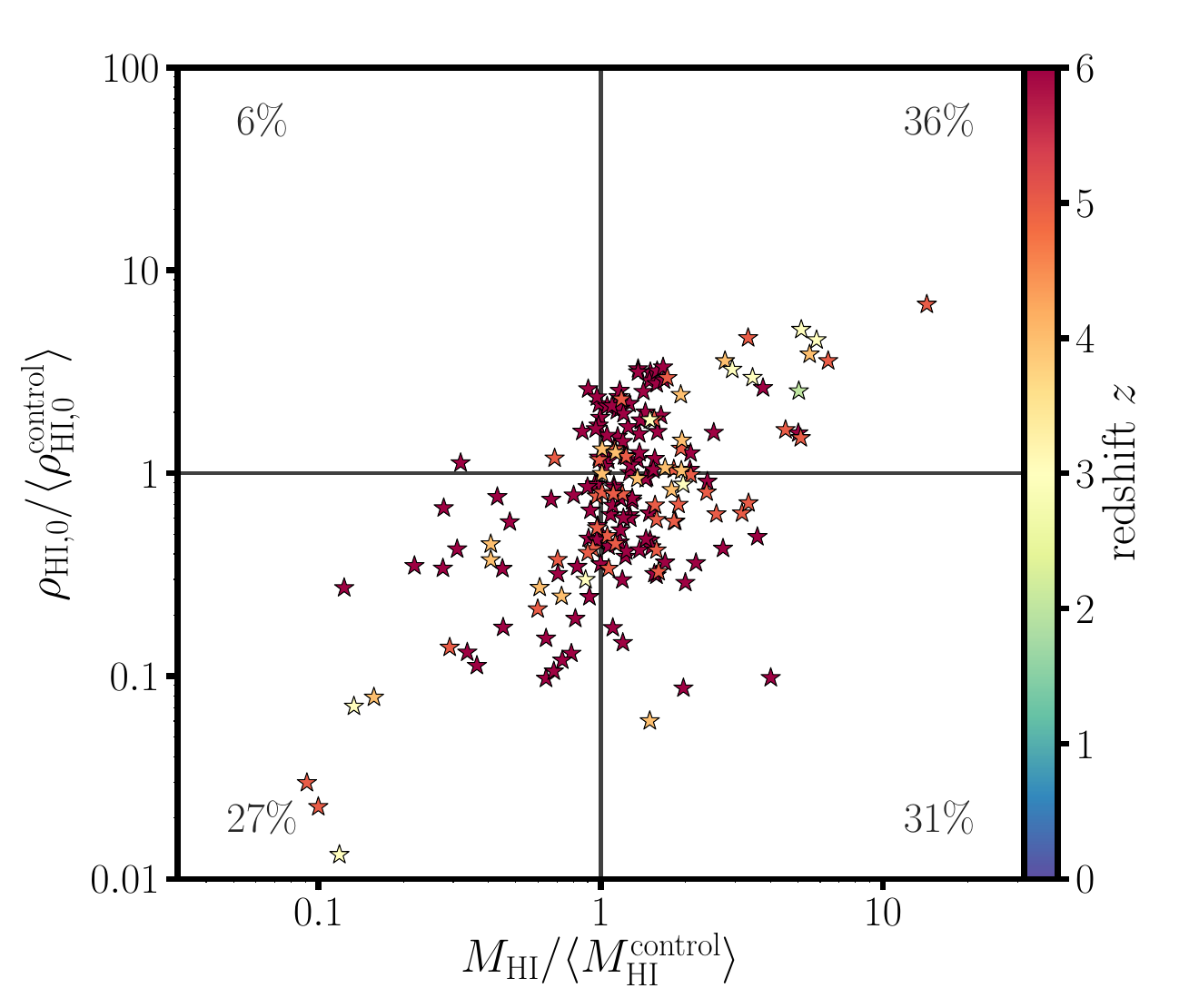}%
\includegraphics[width=0.5\textwidth]{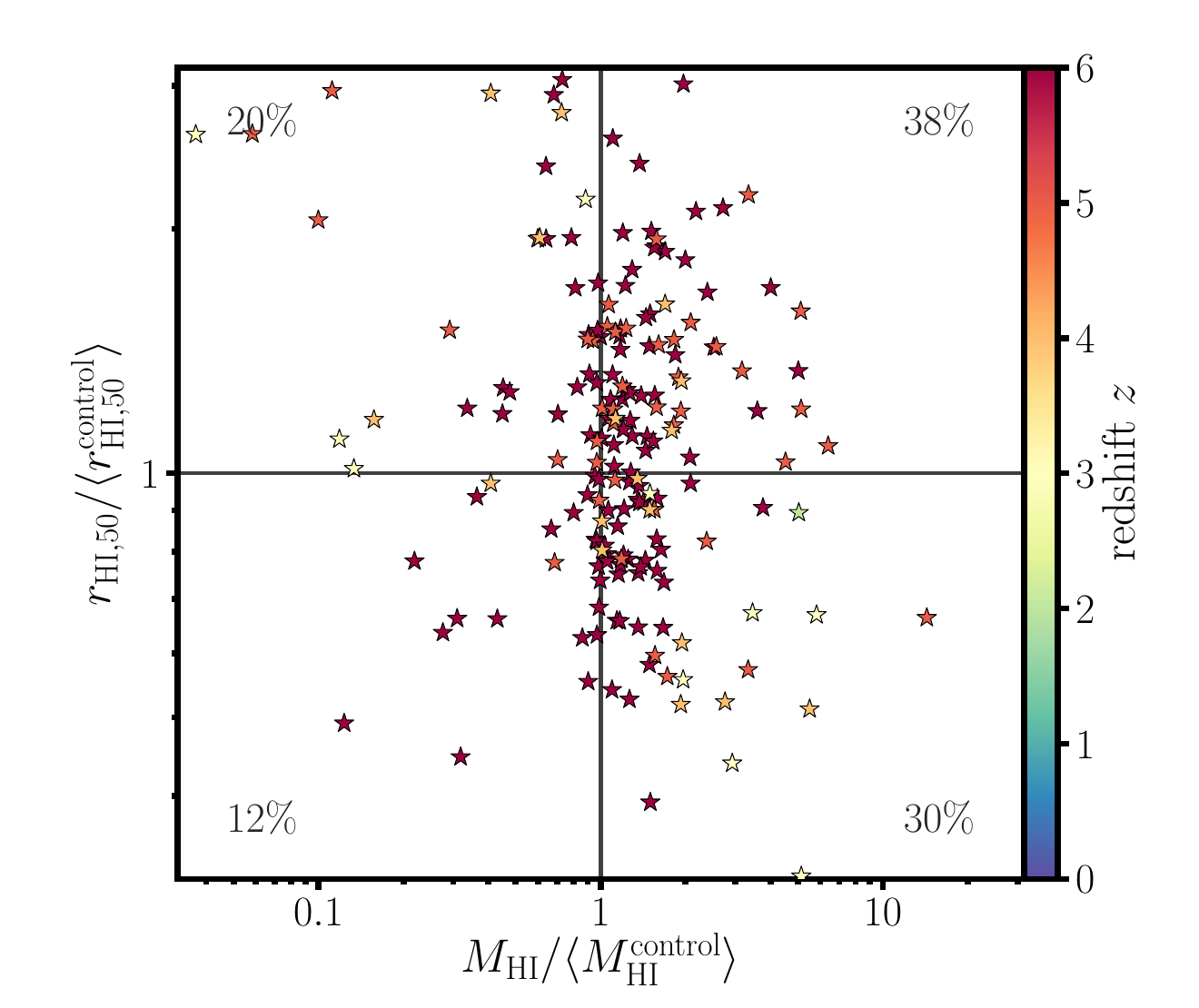}
\caption{The HI reservoir in two planes, raw (\emph{top}) and in enhancement space relative to
$\Mvir$- and $\Mgas$-matched controls (\emph{bottom}). \emph{Left column:} Content versus central
concentration, $\MHI$--$\rhoHIo$. \emph{Right column:} Content versus radial extent,
$\MHI$--$r_{\rm HI,50}/\Rvir$. In the top row, nested gray contours, shaded darker toward higher density, enclose $20$--$99\%$ of all gas-resolved
starless halos ($N=20{,}234$, $z=0$--6). Colored coded curves show the running median of the starless
population in each snapshot, and the black dashed curve represents the pooled $z=2$--$6$ starless median. I.e., the
baseline from which $\Delta\log\rhoHIo$ is measured. Star-shaped symbols represent recently-ignited halos on the same redshift scale (color bars). The $\rhoHIo$ sequence in the left column retreats to lower $\MHI$ with cosmic time, leaving the
region where ignition occurs unpopulated below $z\simeq2$. The right column runs the opposite way. Here, the
pooled starless median descends, so that halos holding more neutral gas hold it over a
\emph{smaller} fraction of $\Rvir$. 
Bottom-row quadrant
percentages display the share of ignited halos on each side of their controls. Because sizes evolve
strongly, controls are drawn from the same snapshot, so the comparison is matched in redshift as well as
in halo and gas mass. \label{fig:plane}}
\end{figure*}

The top-left panel of Figure~\ref{fig:plane} shows that starless halos occupy a tight sequence in the
$\MHI$--$\rhoHIo$ plane (\citet{Spearman1904} rank correlation
 $+0.96$), while ignited halos scatter off (or \emph{decouple} from) this sequence
(Spearman $+0.57$). Namely, among starless halos, knowing $\MHI$ effectively
fixes $\rhoHIo$. However, for ignited halos, the central density has stopped being predictable from its total neutral gas content. It is this \emph{decoupling}, rather than a systematic offset in $\rhoHIo$, that
allows the $\MHI$--$\rhoHIo$ plane to discriminate ignited from starless halos more sharply than $\MHI$ alone.

The top-right panel shows radial extent versus HI content. At fixed halo and gas
mass, recently-ignited halos spread their neutral gas over a \emph{larger} half-mass radius than their
matched-control counterparts (median $r_{\rm HI,50}/\Rvir=0.28$ versus $0.24$). This radial extent depends on the age of the first recently formed star particle: for the youngest half of the sample (age below $21.4$~Myr) the median enhancement is $0.84$ and only $34\%$ of halos exceed their controls, against $1.47$ and $81\%$ for the older half, and the age and the enhancement correlate with a Spearman coefficient of $+0.68$ at fixed redshift. We note that this is not a restatement of their larger $\MHI$. The pooled starless anticorrelation visible in Figure~\ref{fig:plane} (radial
extent decreasing with neutral mass, Spearman $-0.70$) is largely an artifact of stacking epochs of
systematically different neutral mass, and does not hold within individual epochs. This radial-extent
result therefore relies on the matched-control comparison below, not on the raw pooled relation.

The bottom row recasts both planes in enhancement space, relative to their matched controls. The bottom-left
panel shows that ignited halos are preferentially enhanced in $\MHI$ ($67\%$ above their controls),
whereas their central HI density is not ($41\%$). The bottom-right panel shows the same for radial
extent. Comparing each halo against its own matched controls, at its own redshift, $58\%$ of ignited
halos lie above the mean of their own controls. In this matched-control comparison, the independence
between content and radial extent is clear. Namely, $\Delta\log\MHI$ and $\Delta\log (r_{\rm HI,50}/\Rvir)$ are uncorrelated to
within the noise (Spearman $-0.12$, $p=0.11$), against $+0.96$ for $\MHI$ and $\rhoHIo$ among starless
halos. HI mass content and radial extent are thus close to orthogonal descriptions of the same reservoir,
which is why they combine so effectively. In other words, the intrinsic signal tracks the reservoir's
\emph{amount} and its \emph{size}, with $\rhoHIo$ playing a weaker third role.

\subsection{Ranking single features} \label{subsec:singles}
Sections~\ref{subsec:geometric}--\ref{subsec:intrinsic} established how each property behaves on its own. We now ask how they behave together: whether the three families act independently, and what each contributes once the
others are held fixed.

\begin{figure*}[!tbp]
\includegraphics[width=\textwidth]{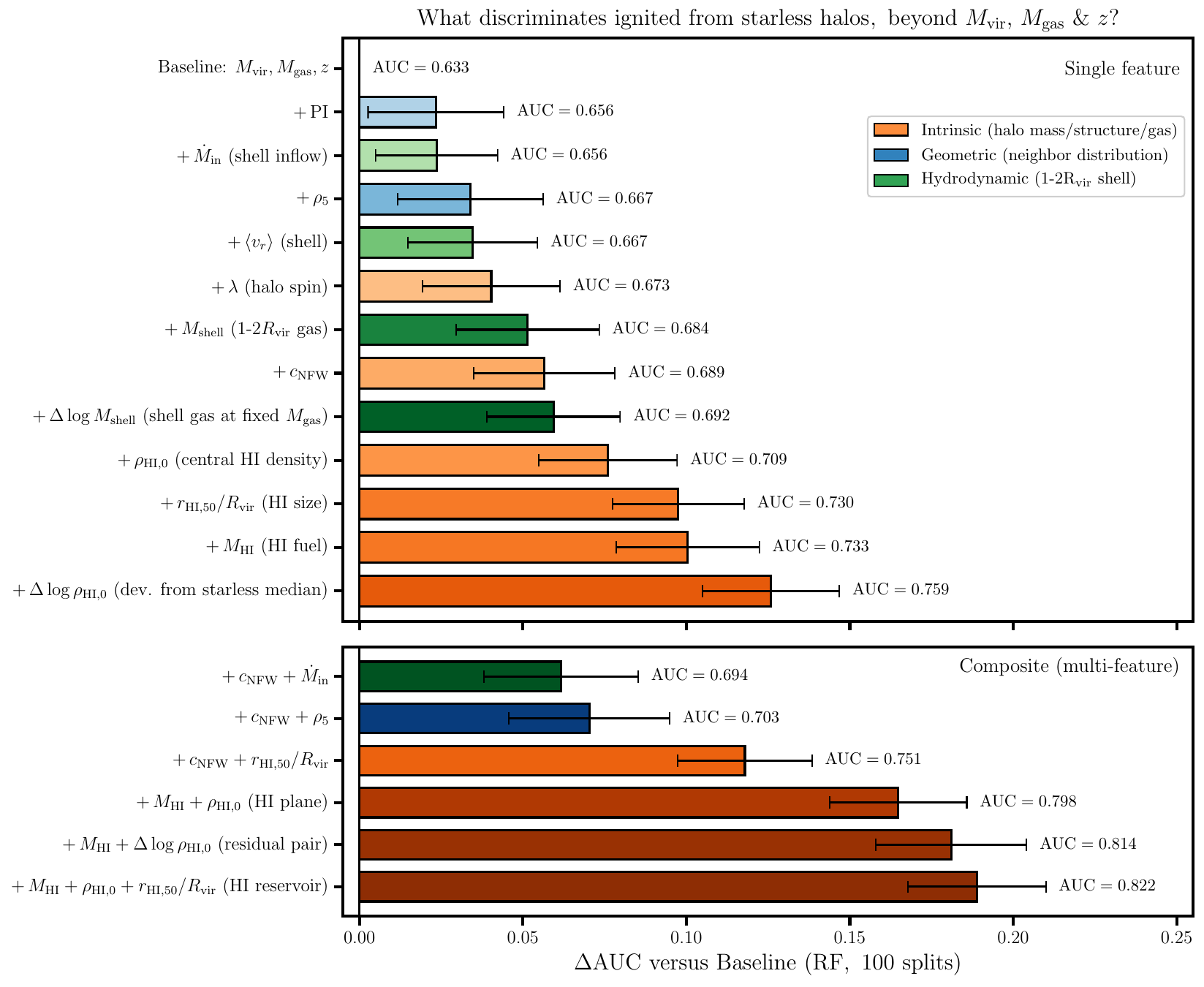}
\caption{Nested-model random-forest test AUC (100 random stratified splits, 30\% train / 70\% test) for each candidate axis added
to the $\Mvir,\Mgas,z$ baseline (0.633), color coded by family (geometric, hydrodynamic, intrinsic),
with darker shades marking higher AUC within each family.
Geometric environment adds little; the intrinsic HI reservoir contributes most, led jointly by the HI
amount $\MHI$ (0.733) and half-mass radius $r_{\rm HI,50}/\Rvir$ (0.730), with the full reservoir
($\MHI+\rhoHIo+r_{\rm HI,50}/\Rvir$) reaching 0.822. Models are split by panel: single added features with the baseline (\emph{top}) and composites of two
or more added features (\emph{bottom}), so that models of the same kind (single versus composite) are compared against one another.
Within each panel, models are ordered by AUC, highest at the bottom. Error bars are the standard deviation of the test AUC across the $100$ splits. Namely, how
far a score moves when the train/test partition is redrawn on the same halos. We note that these error bars do not carry the
separate uncertainty of having only $192$ ignited halos, which is estimated in
Section~\ref{subsec:synthesis}. For reference, every quantity (feature) is defined in Table~\ref{tab:quantities}. \label{fig:regression}}
\end{figure*}

Recall that AUC (Section~\ref{subsec:regression}) runs from $0.5$ (chance) to $1$ (perfect discrimination). The top panel of Figure~\ref{fig:regression} ranks the quantities we have explored thus far as single features against the
baseline of virial mass, gas mass, and redshift. The geometric environment measures add
little. Among the raw hydrodynamic shell
quantities, $M_{\rm shell}$ is that family's best member. However, the intrinsic properties contribute the most. Among them, the HI
half-mass radius ($r_{\rm HI,50}/\Rvir$) and the total HI mass ($\MHI$) are the strongest
single features. They are also indistinguishable from each other, differing by far less than the split-to-split
scatter ($\pm0.02$). The central HI density is a weaker third. Since $\Mgas$ is already in the baseline, we also tested the neutral-gas fraction $f_{\rm HI}=\MHI/\Mgas$ as an alternative single feature, with $\MHI$ and $\Mgas$ both restricted to AHF's bound particles for consistency. This quantity underperforms $\MHI$'s own AUC ($0.695$ versus $0.733$), so the raw amount of retained neutral gas is more informative than its fraction of the halo's bound gas. We report $\MHI$ throughout, both for this reason and because $\Mgas$, unlike $\MHI$, is difficult to measure observationally. We note that Figure~\ref{fig:regression} also includes two new single features, halo concentration and spin. We
introduce and discuss both together with their literature context in Section~\ref{subsec:nulls}.

Two of the original reservoir quantities also enter in residual form. The
shell gas mass, measured as a departure from the tight relation it follows with the halo's own gas
mass, $\Delta\log M_{\rm shell}$, reaches $0.692$, the best of the hydrodynamic family, ahead of the
raw $M_{\rm shell}$ itself. The central density is similarly more informative as a
\emph{deviation} from the starless-halo sequence than
as an absolute value. Replacing $\rhoHIo$ by its residual $\Delta\log\rhoHIo$ raises the
single-feature score from $0.709$ to $0.759$, and
the $\MHI$ pair from $0.798$ to $0.814$. However, the resulting improvement is modest. Namely, $\Delta\log\rhoHIo$ remains below both $\MHI$ and $r_{\rm HI,50}/\Rvir$, and using residuals throughout affects the combined score by less than
the split-to-split scatter.

\subsection{Ranking composite models} \label{subsec:synthesis}
The bottom panel of Figure~\ref{fig:regression} shows the result of combining \emph{multiple} features (composite models). The three HI-reservoir quantities (mass, radial extent and central density) are complementary, and combined they reach a score of $0.822$ (the maximum value in our analysis).

Next, we examine the robustness of our ranking scheme using two methods. First, for the HI-reservoir three-member combination, \emph{resampling the test set} yields a $\Delta$AUC with a $95\%$ interval of $[+0.120,+0.262]$, resolved against the best geometric and
hydrodynamic models alike.  Namely, their $95\%$ intervals do not overlap with that of the best geometric model, and only just touch that of the best hydrodynamic model, which is a stronger statement than merely
avoiding overlap with zero $\Delta$AUC. Second, \emph{average precision} --- which asks how often the
halos ranked most likely to ignite actually do, rather than just how well-ordered the full ranking is
--- tells the same story more sharply. Namely, against a \emph{no-skill value} (the score random guessing would achieve)
equal to the ignited fraction ($0.010$; the full $18{,}284$-halo starless population used throughout this nested-regression analysis, not the per-object matched controls of \S\S~\ref{subsec:rotation}--\ref{subsec:intrinsic}), the baseline reaches roughly twice chance. Meanwhile, adding $\rhofive$
barely moves it, and the reservoir reaches eleven times chance.

Across both checks, geometric environment therefore adds little once mass, gas, and redshift are controlled. Rather, the
strongest discriminant of recent ignition is the intrinsic neutral-gas reservoir -- primarily its mass and radial extent,
with the central peak density a weaker third. We note, however, that the reservoir's advantage over environment holds for recently ignited halos, up to an age limit of $200$~Myr, and is lost when halos of any age are admitted (Appendix~\ref{subsec:agewindow}).

\section{Discussion} \label{sec:discussion}
In Section~\ref{sec:results} we found that -- relative to its large-scale environment or immediate vicinity -- the existing HI-reservoir is the strongest discriminator of whether or not a halo has experienced galaxy ignition recently. Here we discuss this multi-scale chain of events in more detail.

\subsection{The ignition chain: an emerging picture} \label{subsec:reframe}

\begin{figure}[!tbp]
\includegraphics[width=\linewidth]{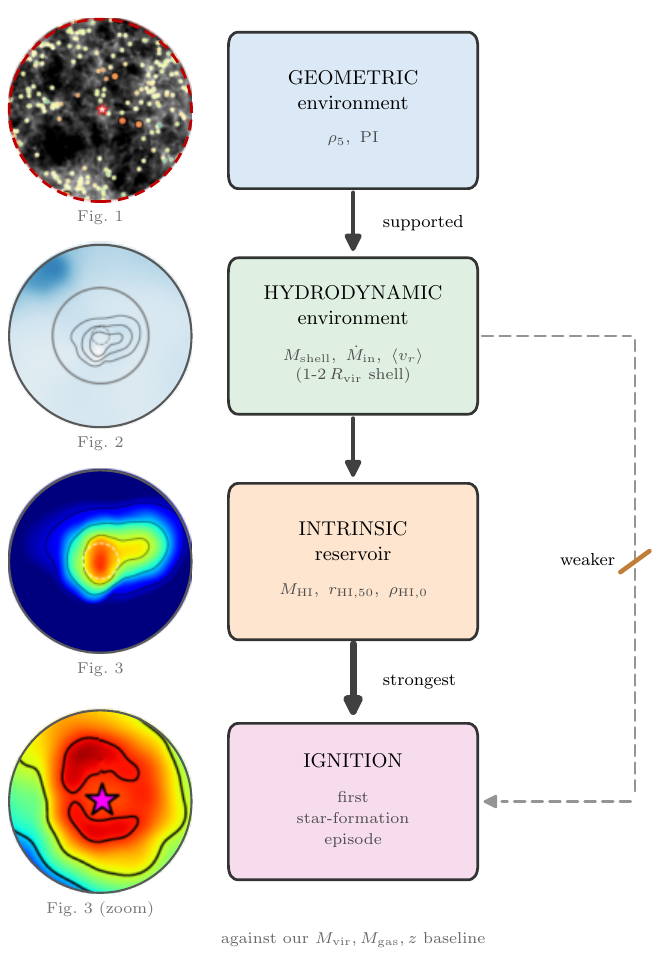}
\caption{The ignition chain. Boxes represent the three families of discriminants
(Section~\ref{subsec:axes}) and the ignition event. Icons are crops of Figures~\ref{fig:web}--\ref{fig:himap} for the representative pair. \emph{Top-to-bottom}: large-scale environment is associated with the gas
supply (the $1-2\Rvir$ shell), which feeds the reservoir, which in turn best discriminates recently ignited halos. The link labeled
\emph{``supported"} indicates that geometric environment is measurably associated with the gas supply. The HI reservoir is the
\emph{``strongest"} step here (it best discriminates ignited halos). Meanwhile, the dashed bypass -- traversing directly from the shell-gas supply to ignition, marked \emph{``weaker"} -- performs worse as a direct discriminator of ignition. This does not rule out an indirect effect through the reservoir: the middle link, from gas supply to reservoir, carries no label, because unlike the other three it is not backed by a regression statistic in this paper. We defer such an investigation to future work.
\label{fig:pipeline}}
\end{figure}

Figure~\ref{fig:pipeline} displays a \emph{chain} consisting of three consecutive stages prior to ignition, each representing one of our three families (large-scale environment, the surrounding $1-2 \Rvir$ shell, and its intrinsic HI reservoir). A mechanism at an early stage must propagate downstream to eventually matter for ignition, and should then leave a measurable signature. It is precisely here where our nested regression scheme (Figure~\ref{fig:regression}) can test this. Below we evaluate each \emph{link} in the chain in order.

Among starless halos, a random-forest regression analysis using only halo mass, total gas mass, and redshift already explains part of the shell's
inflow-rate scatter. Meanwhile, adding the geometric environment measures explains $18$ percentage points more. Moreover, both
correlate positively with inflow, even after controlling for virial mass, total gas mass, and redshift
($+0.37$ for $\rhofive$, $+0.39$ for PI). In other words, denser and more perturbed
large-scale environments do deliver more gas through the shell. Therefore, the \emph{first link} is supported.

However, the \emph{second link} is weak. Namely, the inflow rate is among the weakest discriminants we test,
and the shell family as a whole (green bars  in Figure~\ref{fig:regression}) stays far below the $0.822$ score the halo's own neutral reservoir reaches. Indeed, our results suggest that the \emph{third link} is the strongest discriminator of recent ignition.
That is a stronger claim than simply finding no difference between the populations.\footnote{Figure~\ref{fig:envnull} in Appendix~\ref{sec:appendix} shows the difference to be real and statistically secure. This is in line with Section~\ref{subsec:geometric}, which shows why this difference carries little discriminating power. Namely, only $30\%$ ($26\%$) of ignited halos have higher $\rho_5$ (PI) than their matched controls.}

This claim is consistent with environment acting through \emph{pre-processing}, not geometry: it may set how much raw fuel a halo is offered, but not how the halo utilizes it, and even then, the hydrodynamic axis is secondary to the halo's own neutral-gas reservoir. A natural expectation is that igniting halos should simply be the better-fed ones. However, compared against their own matched controls, we find that they are not. Environment delivers a halo its original endowment of gas, but what best distinguishes recently ignited halos is how the halo transforms that endowment into a reservoir, and ultimately into new stars.

\subsection{The HI reservoir's decoupling}
\label{subsec:whyHI}\label{subsec:colddense}
The reservoir that best discriminates recently ignited halos is not described by a single quantity, but by a composite. At fixed halo mass, gas mass and redshift, an igniting halo holds more HI than its matched controls and, when its first recently formed star particle is older, spreads it over a larger ($\Rvir$-rescaled) half-mass radius. This physical interpretation is contained in the decoupling of Figure~\ref{fig:plane}. This is because, among starless halos, the central density tracks its total HI mass almost perfectly. 

Motivated by the dependence of the radial extent on the age of the first recently formed star particle (Section~\ref{subsec:intrinsic}), we speculate that
the first star-formation event is responsible for rearranging the central gas -- potentially explaining the radial-extent differences between the ignited and starless samples. The outflowing double peanut-shaped overdensity of our representative pair (Figures~\ref{fig:rv} and~\ref{fig:himap}), whose star particle is only $12$~Myr old, is consistent with the onset of this rearrangement. We defer a detailed, population-wide exploration of this mechanism to future work.

Dependence on stellar age is our main clue to the mechanism behind the radially extended HI. In the younger half of our sample (first star particle younger than $21.4$~Myr), the HI is more compact than in the matched controls. The larger radius appears only among halos whose first recently formed star particle is older (Section~\ref{subsec:intrinsic}; Appendix~\ref{subsec:ageextent}), whereas the amount of HI is already enhanced in the youngest halos. A more extended reservoir therefore does not precede ignition in our data, it accompanies or follows it. This points to a process activated by the first star-formation event, most plausibly stellar feedback that redistributes the gas to larger radii, rather than to a pre-existing property that favors ignition. Two findings support this interpretation. First, the HI-weighted radial velocities inside $r_{\rm HI,50}$ are positive in ignited halos (Appendix~\ref{subsec:ageextent}). Second, the central density $\rhoHIo$ is not enhanced even though the total HI mass is, as expected if the additional gas sits at larger radii. In this picture the extent discriminates ignited from starless halos in part because it records the response of the gas to the first star. However,  its discriminating power need not reflect a role in triggering ignition.

\subsection{Evolution: the closing of the ignition~window} \label{subsec:evolution}
\begin{figure*}[!tbp]
\includegraphics[width=\textwidth]{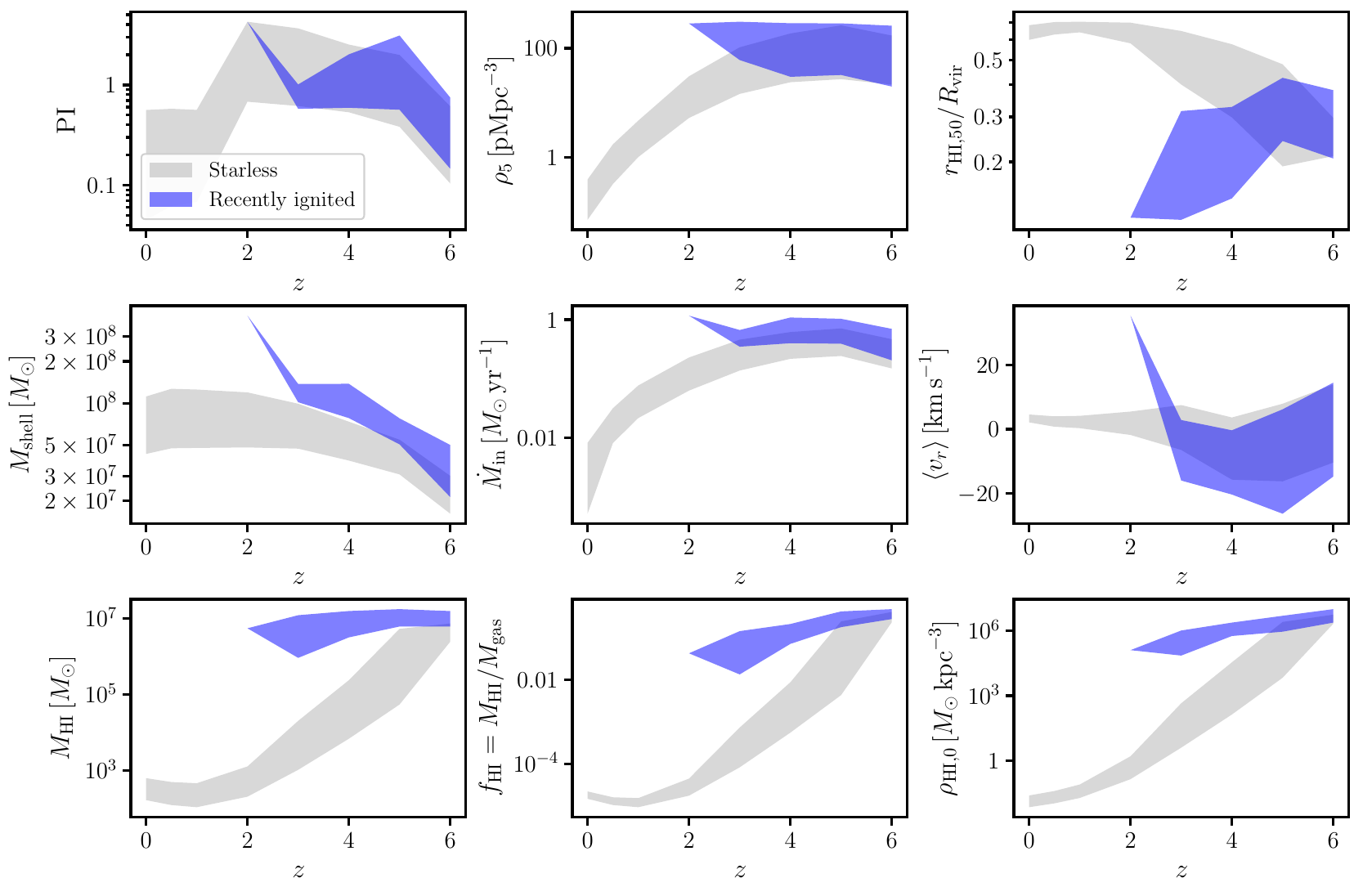}
\caption{Redshift evolution (running median with 25--75 percentile bands) of the geometric
(\emph{top}), hydrodynamic (\emph{middle}), and intrinsic (\emph{bottom}) axis quantities, for gas-resolved
starless halos (gray, $z=0$--6) and recently-ignited halos (blue, $z=2$--6). The bottom row is ordered as
amount, ionization state, and concentration of the reservoir. The neutral fraction
$f_{\rm HI}=\MHI/\Mgas$ tracks the HI decline, falling by $4.3$~dex
(Section~\ref{subsec:evolution}). Ignited halos remain roughly two decades more neutral than the starless
population at every epoch they exist. Population offsets here are not matched in $\Mvir$ or $\Mgas$ and
several reverse under matching (Section~\ref{subsec:hydro}); the $z=2$ ignited bands rest on a single
object. \label{fig:evolution}}
\end{figure*}

Above we ask what distinguishes an igniting halo from a starless one \emph{at a given epoch}. The complementary question is how these populations change with redshift, which we explore below (Figure~\ref{fig:evolution}). In particular, we discuss why the igniting population
vanishes below $z\simeq2$ \citep{Moreno2026}.

Recall that Figure~\ref{fig:plane} shows the evolution of the starless $\MHI$--$\rhoHIo$ sequence, which experiences a mild shift downwards with decreasing redshift.  However, the main evolutionary effect here is that this relation \emph{retreats} to progressively lower neutral gas masses, away from the region the ignited halos occupy. Overall, recently-ignited halos have $\log\MHI=7.0$ at the median, and $90\%$ of them exceed $\log\MHI=6.2$. The
fraction of starless halos reaching this lower level falls steeply with cosmic time, from $89\%$ at $z=6$ to below
$1\%$ by $z=2$, where recently-ignited halos stop appearing altogether.

The reason behind this retreat is not that these halo populations lose their gas. Between $z=6$ and $z=0$, the median starless halo grows
by $1.2$~dex in $\Mvir$, while its gas mass is unchanged ($+0.05$~dex). Rather, accretion is suppressed, so the
gas fraction falls by more than an order of magnitude simply because the halo grows and the gas does not \citep[Figure~4 of][]{Moreno2026}.
What plummets is the \emph{neutral} gas fraction $f_{\rm HI}$ (Figure~\ref{fig:evolution}, bottom
center), by more than four decades, from $\MHI/\Mgas=0.32$ at $z=6$ to effectively zero below $z\simeq2$. In other words, the reservoir is not
drained but ionized. In sum, two effects compound here. Ionization by the UV background removes the neutral gas, and a halo that cannot accrete efficiently cannot replace it. Ionization accounts for nearly all of the decline in HI, because the gas mass is unchanged while the neutral fraction falls by more than four decades. The hydrodynamic panels of Figure~\ref{fig:evolution} show the second effect. For the median starless halo, the shell gas mass stops growing below $z\simeq3$. The inflow rate through the shell falls from $0.41\,M_\odot\,{\rm yr}^{-1}$ at $z=5$ to $0.0023\,M_\odot\,{\rm yr}^{-1}$ at $z=0$, and the mean radial velocity changes from net inflow ($-3.2\,{\rm km\,s^{-1}}$ at $z=5$) to a small net outflow ($+3.5\,{\rm km\,s^{-1}}$ at $z=0$). The supply of fresh gas therefore dries up, so the neutral gas removed by ionization is no longer replaced. These trends are consistent with the closing of the ignition window, although this comparison of entire populations does not show how much each effect contributes.

Nevertheless, the intrinsic HI reservoir is not the only property that evolves. Figure~\ref{fig:evolution} traces our three families as running medians with interquartile bands. Note that this
is a comparison of \emph{entire populations}, not of matched-pair samples.\footnote{For a discussion on the disadvantages of only comparing entire populations, see \S~3.4 and \cite{Moreno2026}.} Concretely, an offset here can therefore reflect
the mass difference between the two populations, rather than a genuine difference in any one quantity. Furthermore, several quantities -- for instance the perturbation index (Section~\ref{subsec:geometric}), the shell gas mass (Section~\ref{subsec:hydro}), and the HI radial extent (Section~\ref{subsec:intrinsic}) -- flip direction entirely once matched. One quantity that refuses to flip is the total neutral mass. At every redshift, recently-ignited halos harbor more HI gas than their starless counterparts, matched or not. Moreover, this
gap widens as cosmic time erodes the population down to a handful of survivors. Indeed, the signature that
discriminates recently ignited halos at different epochs is not a halo's location within the cosmic web, but its intrinsic HI reservoir.

\subsection{Comparison with previous work} \label{subsec:nulls}
Compared to Paper~I, which establishes both that halos ignite in FIREbox and at what time after the end of reionization, here we quantify what
distinguishes the halos that ignite. Moreover, we also add the environmental axes not previously tested in that work. Indeed, the nature-versus-nurture and assembly-bias literature \citep{Zu2018} finds that intrinsic
halo properties matter more than environment for galaxy colors at $z\sim0$. This notion applies here as well: intrinsic properties are better discriminants of galaxy ignition than environment.

We now return to the literature context mentioned in Figure~\ref{fig:regression} (spin and concentration). Recent Local-Group simulations by \citet{GarciaBethencourt2026} find that HI-bearing dark galaxies at $z=0$
occupy less-concentrated, higher-spin halos, motivating us to test spin here \citep{Bullock2001}. Likewise, \citet{Lee2024} report higher spins for dark galaxies at $z=0$ in
\textsc{tng50} \citep{TNG50} and a separation that widens with time. Matched against their own controls, we find that starless halos spin \emph{faster} than their igniting counterparts (median $\lambda=0.056$ against
$0.049$). Moreover, we find that only $36\%$ of ignited halos exceed their own controls ($p=3\times10^{-7}$). We highlight that although the
offset is measurable, it is weak as a discriminant, adding only $0.040$ in AUC over the baseline.

So far we have only found our results to be consistent with the spin portion of the \citet{GarciaBethencourt2026} picture. 
Regarding concentration, this quantity reaches ${\rm AUC}=0.689$ as a single feature, below every HI reservoir
quantity and comparable with the hydrodynamic environment ones. This signal survives the matching on $\Mvir$, $\Mgas$ and $z$. At fixed halo mass, concentration correlates with assembly history. Namely, halos that assemble earlier tend to be more concentrated \citep{Wechsler2002,Ludlow2014}, so a residual signal is what one expects if earlier-forming halos, with deeper central potentials, are better able to retain gas against the ionizing background and to compress and cool it. This is the picture proposed in Paper~I, where recently ignited halos tended to be more concentrated and richer in cold dense gas than their matched starless counterparts \citep{Moreno2026}. It is also consistent with \citet{Fitts2017}, who report that low concentration might be the culprit behind halos failing to form stars, and with \citet{Sawala2016}, whose luminous halos form earlier and are more concentrated than their dark counterparts. We stress what the present test can and cannot show. Our regression contains no direct measure of assembly history (formation time or merger history), so we cannot attribute the signal to assembly rather than to spin, environment or another quantity correlated with concentration, all of which Paper~I listed among the possible origins of high concentration. This estimator is also sensitive to how well the rotation curve peak is resolved (Appendix~\ref{app:quantities}), so we regard the signal as supporting this picture, not as demonstrating it.

Beyond halo structure, a different hypothesis is that environment suppresses ignition, rather
than merely failing to trigger it. Field low-mass galaxies
can be quenched by the cosmic-web
\citep{Benitez-Llambay2017,Luber2025}, or by a backsplash passage through a massive neighbor's circumgalactic
medium \citep{Benavides2021,Moreno2022,Benavides2025,Bhattacharyya2025}. The mechanism that removes the gas in either case is ram pressure \citep{Abadi1999,Samuel2022,Samuel2023}. Such a channel would
affect any gas-resolved halo in unforeseeable ways.

By requiring a resolved gas component, our sample already pre-selects for cosmic-web and backsplash
stripping survivors. This paper's finding that intrinsic properties discriminate better than environment therefore applies to halos that have already retained enough gas to be resolved; it cannot determine whether environment instead influences that earlier retention stage, where its effect may be strongest.

\section{Conclusions} \label{sec:conclusions}
We compare recently-ignited halos with mass- and gas-matched gas-resolved starless halos in \texttt{FIREbox}
across $z=2$--$6$, by ranking geometric, hydrodynamic, and intrinsic properties as discriminants of galaxy ignition. Each
property is scored by its gain in AUC (see Section~\ref{subsec:regression} for machine-learning details) over a baseline of virial
mass, gas mass and redshift, which alone reaches $0.633$ (Figure~\ref{fig:regression}). Our main
conclusions are as follows:
\begin{itemize}
\setlength{\itemsep}{0.02cm}
\setlength{\parsep}{0pt plus 0pt minus 0pt}
\setlength{\topsep}{0pt plus 0pt minus 0pt}
\item \textbf{Large-scale \emph{geometric} environment (the perturbation index, PI, and the local density, $\rhofive$)
carries almost no discriminating power for ignition} once virial mass, gas mass, and redshift are fixed. 
PI and $\rhofive$ reach ${\rm AUC}=0.656$ and $0.667$. This is
not because the two populations share an environment. Rather, against matched controls, recently ignited halos are
consistently displaced toward \emph{poorer} surroundings (by $0.46$~dex in $\rhofive$, enhanced in only $30\%$ of pairs). But the environmental
distributions are broad, so even that displacement separates the populations weakly.
\item \textbf{\emph{Hydrodynamic} environment, the gas supply through the $1$--$2\,\Rvir$ shell, is
modestly informative}. Its strongest member, the shell gas mass measured at fixed halo gas mass,
reaches an AUC of $0.692$, ahead of the geometric environmental measures, but below the intrinsic properties.
\item \textbf{Environment is nonetheless not inert.} It is measurably associated with that gas supply. Namely, adding the geometric measures to a baseline of mass, gas mass and redshift raises the explained variance in the shell inflow rate from
$34\%$ to $52\%$, with partial rank correlations of $+0.37$ and $+0.39$ at fixed mass, gas mass and redshift. Thus, the link from environment to gas supply is real, while the bypass from gas supply to ignition is a weaker discriminator. In other words, environment is associated with the delivery of fuel but not with its use.
\item \textbf{The \emph{intrinsic} HI reservoir is the strongest discriminator of recent ignition along two nearly-equal axes}. These are the \emph{amount} of neutral gas ($\MHI$, $0.733$) and its \emph{radial extent} (the half-mass radius
$r_{\rm HI,50}/\Rvir$, $0.730$). The central density $\rhoHIo$ is a weaker third ($0.709$). Combined, the
three reach $0.822$, the \emph{highest} AUC score we measure in our nested-regression analysis.
\item \textbf{At fixed halo mass, gas mass and redshift, recently-ignited halos hold more HI, and spread it over a radius that is larger when the first recently formed star particle is older}. The mass and extent enhancements are uncorrelated (Spearman $-0.12$,
$p=0.11$), so radial extent is not a restatement of mass content. Ignition tracks the HI reservoir a halo has assembled, not where
it sits on the cosmic web. Ignited halos scatter off the
tight $\MHI$--$\rhoHIo$ sequence that their starless counterparts trace.
\item \textbf{The ignition window closes as the HI is ionized and its supply dries up, not by stripping.} \texttt{FIREbox} produces no recently-ignited halos
below $z\simeq2$ \citep{Moreno2026}. This behavior tracks the HI reservoir itself. The median starless halo 
maintains its gas mass across epochs (a difference in $+0.05$~dex from $z=6$ to $z=0$), while its neutral fraction falls by $4.3$~dex. In other words, the HI reservoir
that discriminates recently ignited halos is ionized away rather than stripped. Its supply also dries up: the inflow rate through the $1$--$2\,\Rvir$ shell falls by more than two orders of magnitude from $z=5$ to $z=0$, so the neutral gas removed by ionization is no longer replaced. Likewise, the fraction of starless halos reaching the
neutral mass exceeded by $90\%$ of ignited ones falls from $89\%$ at $z=6$ to below $1\%$ by $z=2$.
\end{itemize}
\vspace{0.15cm}
Together, these results converge on a single conclusion. In FIREbox, a halo's own neutral gas discriminates recently ignited halos better than its environment does. Nurture, a halo's environment,
adds little discriminating power. We use ``nature'' and ``nurture'' descriptively, for intrinsic and environmental properties; a ranking by discriminating power does not by itself establish causation. We plan to examine discrimination versus causation in more detail in future work.

\begin{acknowledgments}
J.M. wishes to thank his mother, Martha Soto Hern\'andez, for hosting him while writing this manuscript. J.M. is funded by a Pomona College Large Research Grant. D.A., E.L., V.W. and E.S. were supported by Pomona College Summer Undergraduate Research Program (SURP) awards. C.W. acknowledges the CPP Provost's Teacher-Scholar Program. F.J.M. was supported by National Science Foundation (NSF) Math and Physical Sciences (MPS) Award AST-2316748. M.K.R.W. acknowledges support from NSF MPS Ascending Faculty Catalyst Award AST-2444751. R.F. acknowledges financial support from the Swiss National Science Foundation (grant No. 200021-188552). The authors wish to thank the anonymous reviewer for their insightful comments. The authors also acknowledge the labor by the cleaning and the clerical staff, the food service workers, the technical support personnel, and many others that make the astronomy research enterprise possible. This work was conducted on Tongva--Gabrielino land.
\end{acknowledgments}
\software{yt \citep{Turk2011}, Amiga Halo Finder \citep{Knollmann2009}, scikit-learn \citep{Pedregosa2011}.}

\appendix
\restartappendixnumbering
\section{Reference guide: terminology}
\label{app:quantities}
This appendix gathers the measurement details -- apertures, sign conventions, units -- for every
quantity used in the analysis, summarized in Table~\ref{tab:quantities}.

\begin{deluxetable}{rlr}[!t]
\tabletypesize{\footnotesize}
\tablecaption{Every quantity entered into our nested-regression analysis (Figure~\ref{fig:regression}),
grouped by family. \label{tab:quantities}}
\tablehead{\multicolumn{1}{l}{Symbol} & \colhead{Quantity} & \multicolumn{1}{r}{Defined in}}
\startdata
\multicolumn{3}{l}{\textbf{Baseline}} \\
$\Mvir$ & Virial mass & \S\ref{subsec:samples} \\
$\Mgas$ & Gas mass within $\Rvir$ & \S\ref{subsec:samples} \\
$z$ & Redshift & \S\ref{subsec:samples} \\
\multicolumn{3}{l}{\textbf{Geometric environment}} \\
PI & Perturbation index & \S\ref{subsec:axes} \\
$\rhofive$ & Local galaxy number density & \S\ref{subsec:axes} \\
\multicolumn{3}{l}{\textbf{Hydrodynamic environment}} \\
$M_{\rm shell}$ & Gas mass in the $1$--$2\,\Rvir$ shell & \S\ref{subsec:axes} \\
$\Delta\log M_{\rm shell}$ & Shell gas mass at fixed $\Mgas$ & \S\ref{subsec:axes} \\
$\dot M_{\rm in}$ & Inflow rate through the shell & \S\ref{subsec:axes} \\
$\avgvr$ & Mean radial velocity in the shell & \S\ref{subsec:axes} \\
\multicolumn{3}{l}{\textbf{Intrinsic: neutral reservoir}} \\
$\MHI$ & Total neutral-hydrogen mass & \S\ref{subsec:axes} \\
$r_{\rm HI,50}/\Rvir$ & HI half-mass radius, in units of $\Rvir$ & \S\ref{subsec:axes} \\
$\rhoHIo$ & Central HI density within $0.15\,\Rvir$ & \S\ref{subsec:axes} \\
$\Delta\log\rhoHIo$ & Central HI density at fixed $\MHI$ & \S\ref{subsec:axes} \\
\multicolumn{3}{l}{\textbf{Intrinsic: halo structure}} \\
$\cNFW$ & Concentration & \S\ref{subsec:axes} \\
$\lambda$ & Halo spin & \S\ref{subsec:nulls} \\
\enddata
\end{deluxetable}

\textbf{Baseline.} The virial mass $\Mvir$ and radius $\Rvir$ are \textsc{ahf}'s
(Section~\ref{subsec:samples}), $\Mgas$ is the gas mass within $\Rvir$, and $z$ is the redshift at which the halo is measured. The virial velocity, used below to define concentration and spin,
is $V_{\rm vir}=(G\Mvir/\Rvir)^{1/2}$.

\textbf{Geometric environment.} Both measures are built from the galaxy catalog of the same snapshot: the
perturbation index PI sums the tidal influence of the surrounding galaxies, and $\rhofive$ is the number
density within the sphere reaching the fifth-nearest neighbor. Section~\ref{subsec:axes} provides both
expressions. 

\textbf{Hydrodynamic environment.} The following four quantities are measured within the spherical shell between $1$
and $2\,\Rvir$, from all gas particles in that shell (we extract the full $2\,\Rvir$ sphere of every halo, so that gas whose smoothing kernel overlaps the shell is not truncated). Radial velocities are mass-weighted, taken with respect to the halo's bulk motion, and
include the Hubble flow across the shell, so $\avgvr<0$ denotes net inflow. The inflow rate is defined as the mass
flux carried by inward-moving gas, $\dot M_{\rm in}=\frac{1}{\Delta R}\bigl|\sum_{v_{r,i}<0} m_i\,v_{r,i}\bigr|$,
with shell width $\Delta R=2\Rvir-\Rvir = \Rvir$.

\textbf{Intrinsic: halo structure.} The concentration $\cNFW$ is \textsc{ahf}'s, in the
\citet{Prada2012} form inferred from $V_{\rm max}/V_{\rm vir}$ rather than from a profile fit
(Section~\ref{subsec:axes}). This estimator is sensitive to how well the rotation curve peak is
resolved. The spin $\lambda$ is \textsc{ahf}'s \citet{Bullock2001} parameter,
$\lambda=J/(\sqrt{2}\,\Mvir V_{\rm vir}\Rvir)$.%

\textbf{Intrinsic: neutral reservoir.} The three reservoir quantities are computed relative to the HI density peak. We locate the peak with an iterative shrinking-sphere centroid, restricted to gas within $\Rvir$ because a halo with a neutral companion just outside its virial radius can otherwise have the shrinking sphere captured by that object. $r_{\rm HI,50}$ is the radius from that peak enclosing half the within-$\Rvir$ HI mass, and $\rhoHIo$ is the HI mass within $0.15\,\Rvir$ of the peak divided by the enclosed volume. We adopt an aperture of $0.15\,\Rvir$ following Paper~I \citep[see also][]{Wheeler2025,Kravtsov2013}. We quote the HI extent in units of $\Rvir$ rather than in kpc because physical HI sizes evolve steeply with $z$ while their ratio to $\Rvir$ does not (Section~\ref{subsec:intrinsic}).

\vspace{-0.1cm}
\section{Distributions: three families}
\label{sec:appendix}
\begin{figure*}[tbp]
\includegraphics[width=0.5\textwidth]{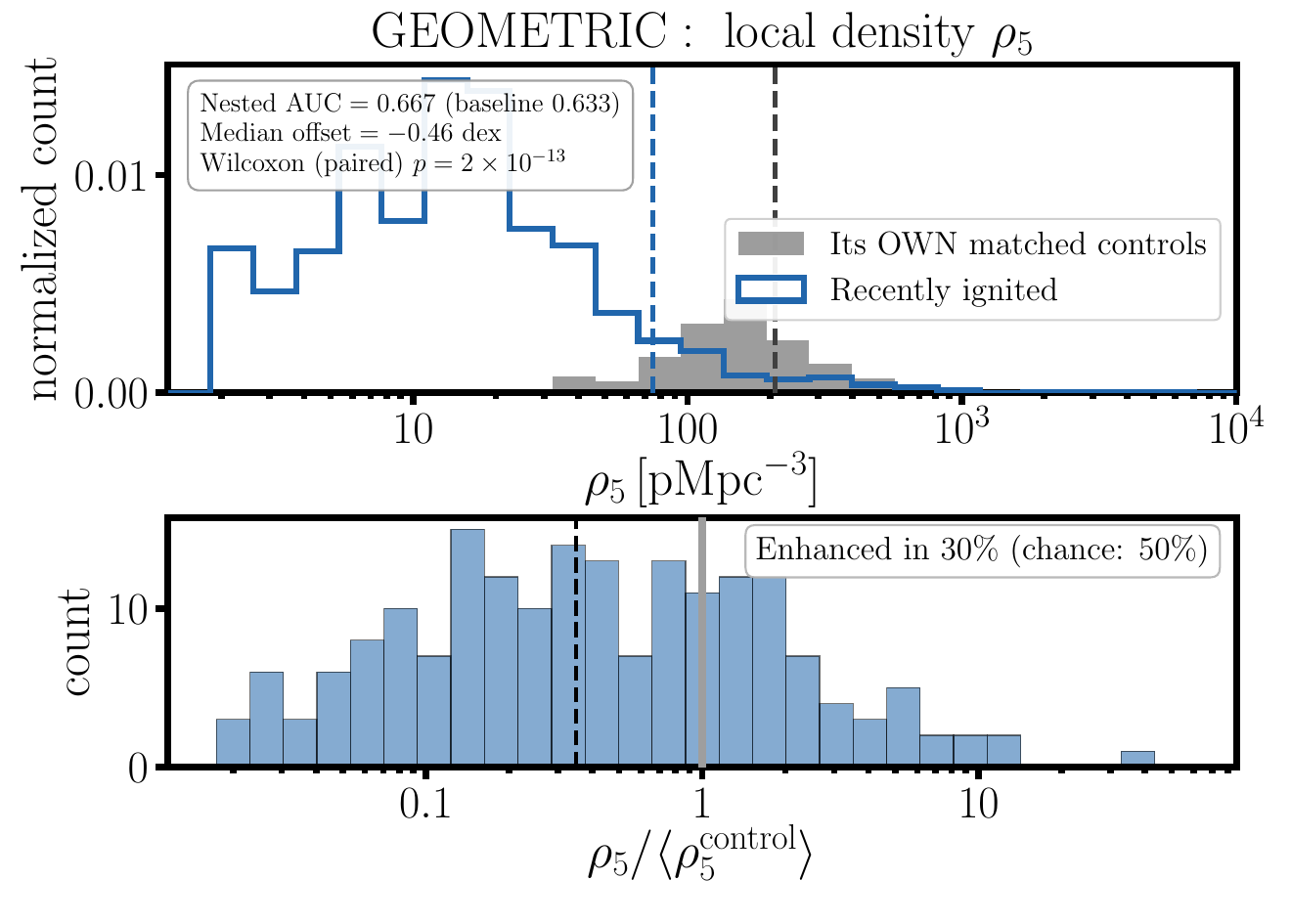}%
\includegraphics[width=0.5\textwidth]{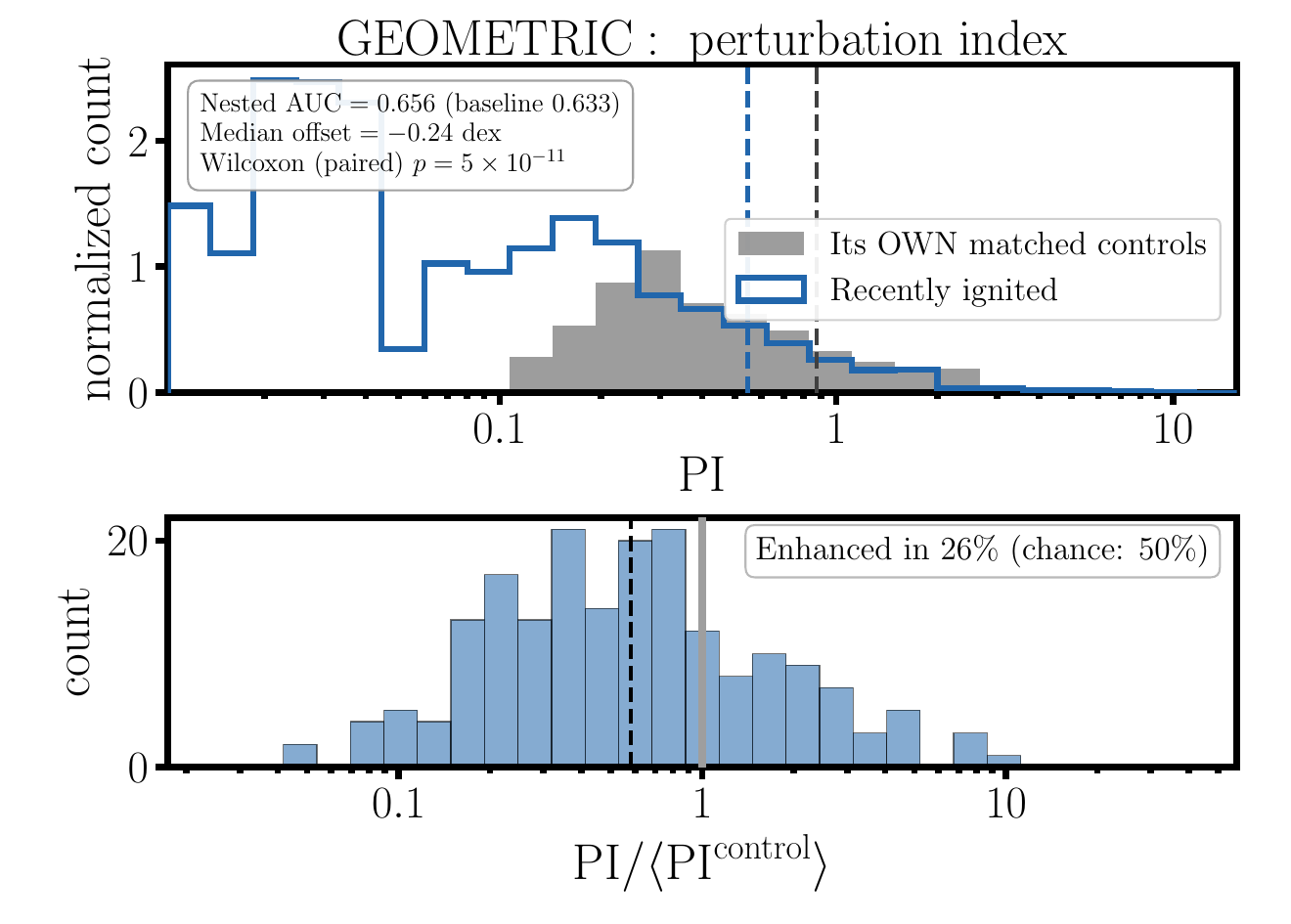}\\[2pt]
\includegraphics[width=0.5\textwidth]{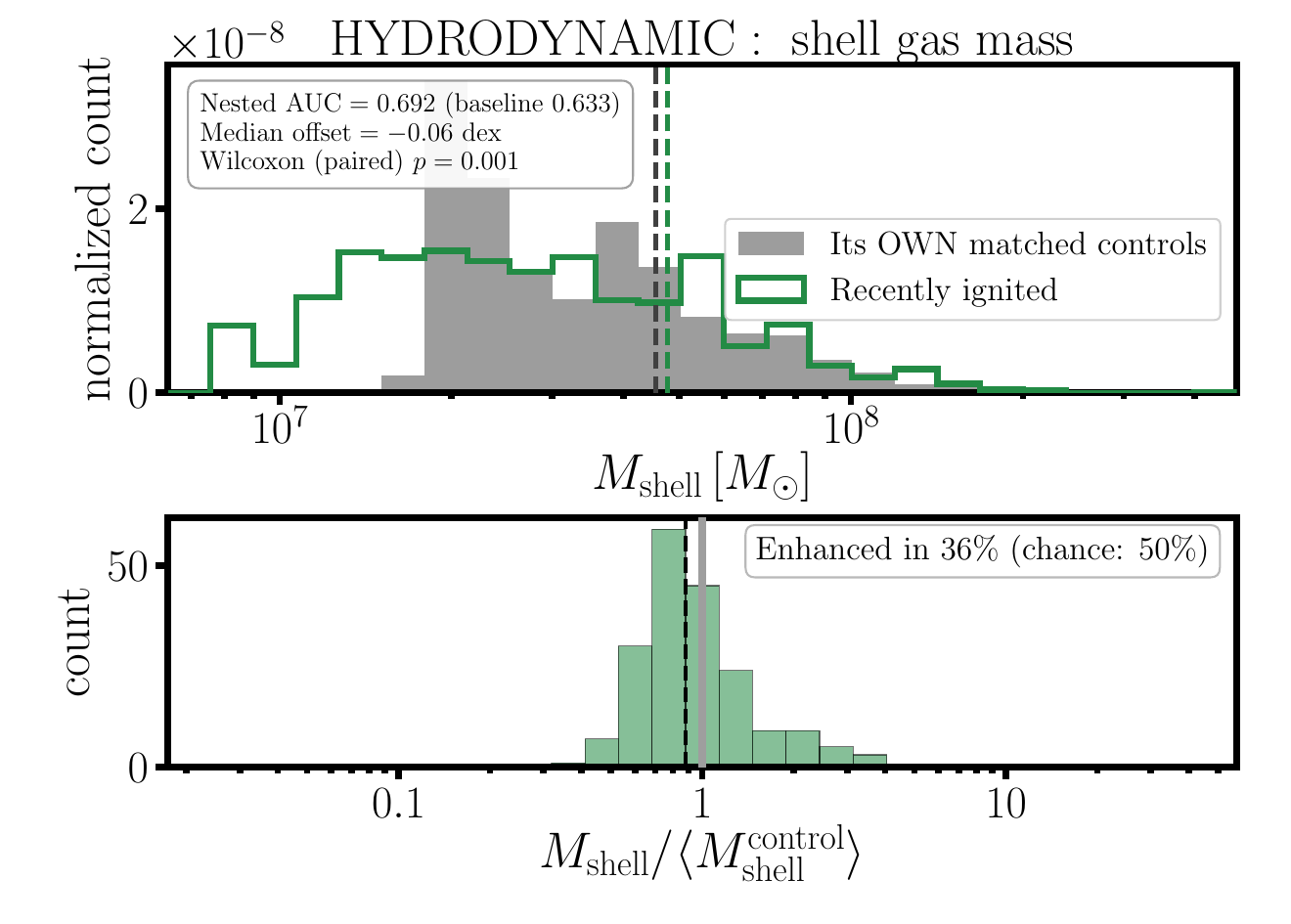}%
\includegraphics[width=0.5\textwidth]{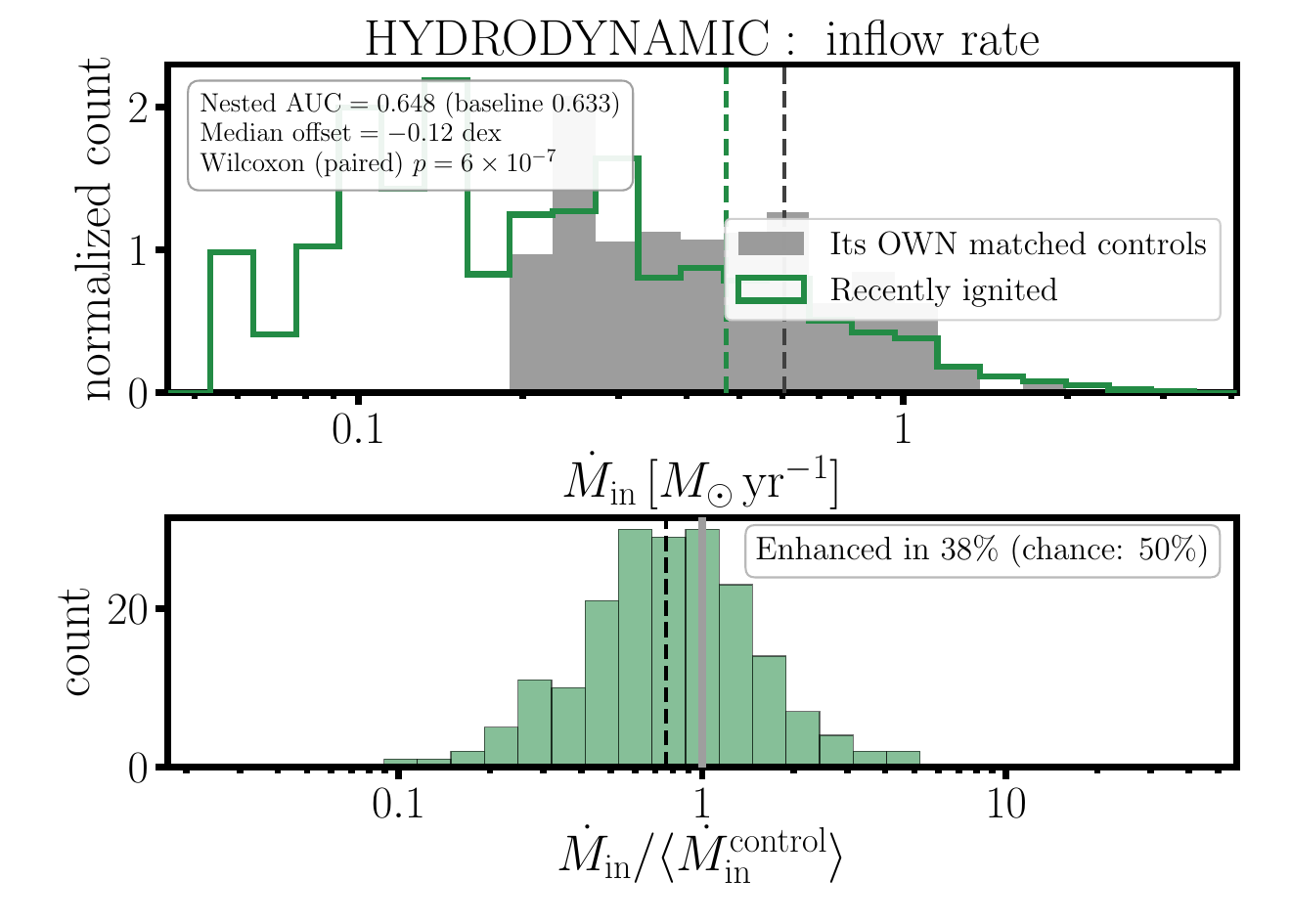}\\[2pt]
\includegraphics[width=0.5\textwidth]{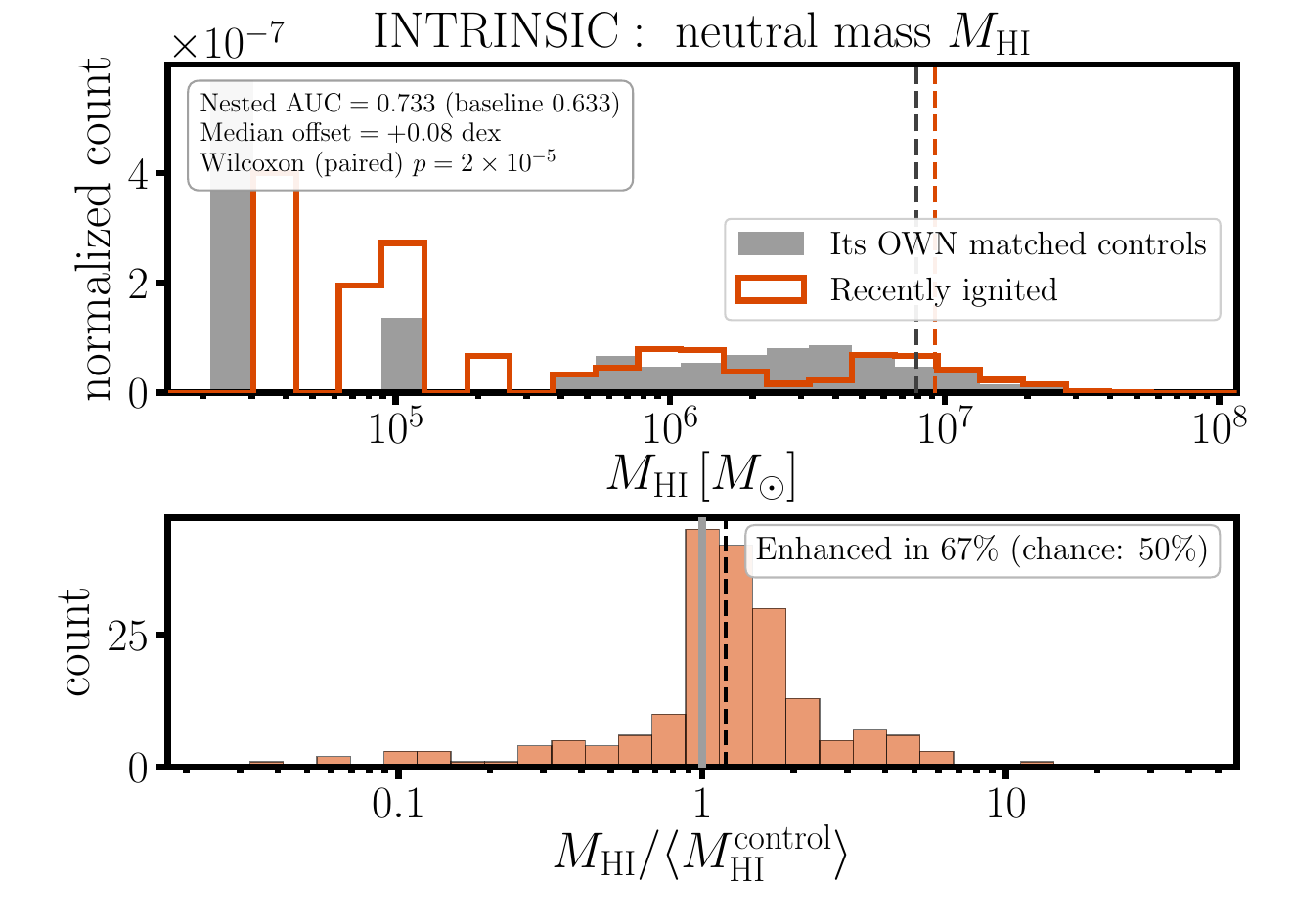}%
\includegraphics[width=0.5\textwidth]{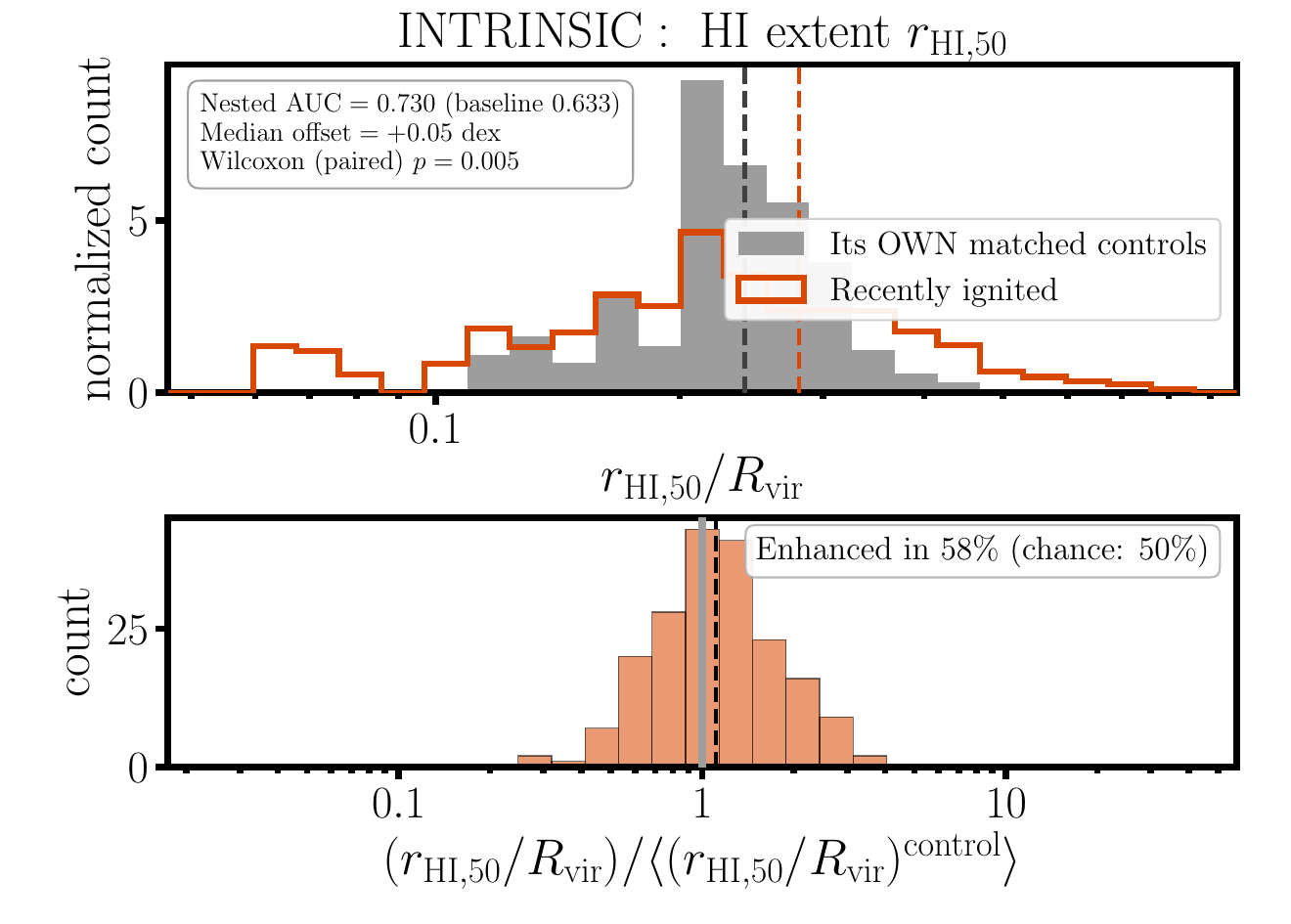}
\caption{Population statistics for the three families. We report each quantity for recently-ignited halos against
\emph{their own} matched controls ($N=192$ ignited halos, each with its own control set). A halo's
controls are the starless halos in the same redshift whose virial and gas masses match it to within
$0.1$~dex. Rows are geometric (\emph{top}: $\rhofive$, PI) and hydrodynamic
(\emph{middle}: $M_{\rm shell}$, $\dot M_{\rm in}$) environment, plus intrinsic (\emph{bottom}: $\MHI$,
$r_{\rm HI,50}/\Rvir$) properties. Panels are colored coded by family, matching Figure~\ref{fig:regression}: blue for geometric, green
for hydrodynamic, orange for intrinsic. Upper histogram: The ignited distribution (color coded outline)
against the starless halos matched to them individually in $\Mvir$, $\Mgas$ and redshift (gray),
medians dashed. Lower histogram: The paired
enhancement, one entry per ignited halo, median dashed and unity
solid. Each panel quotes the nested-regression AUC against the mass--gas--redshift baseline, the median
offset in dex, the paired Wilcoxon signed-rank $p$ \citep{Wilcoxon1945}, and the fraction of pairs
enhanced.
\label{fig:envnull}}
\end{figure*}

Figure~\ref{fig:envnull} sets each quantity against its own matched control set, one row per family
(geometric, hydrodynamic, intrinsic), with the two distributions above and their pairwise ratio
below. It is the distribution-level view behind the summary statistics of Figure~\ref{fig:enhflow},
and it is where the paired significance of each comparison is reported. The distributions make
explicit why a large median offset need not imply strong discriminating power: broad distributions,
as for the environmental quantities, let even a large shift separate the populations poorly, while
narrow ones, as for the reservoir, let a small shift separate them well. Discriminating power follows
the enhancement fraction, not the offset in dex.

\setcounter{figure}{0}  %
\section{Robustness checks}
\label{sec:appC}
\subsection{The recent-ignition window}
\label{subsec:agewindow}
Our recently ignited sample consists of halos with a single star particle younger than $100$~Myr (\S\ref{subsec:samples}). To test whether the ranking of Figure~\ref{fig:regression} depends on that limit, we repeat the nested regression with the same starless class and baseline, defining the ignited class by a stellar age below $25$, $50$, $100$ or $200$~Myr, and with no limit (all $2{,}593$ halos with a single star particle within $0.15\,\Rvir$ at $z=2$--$6$).

\begin{figure}[tbp]
\centering
\includegraphics[width=\columnwidth]{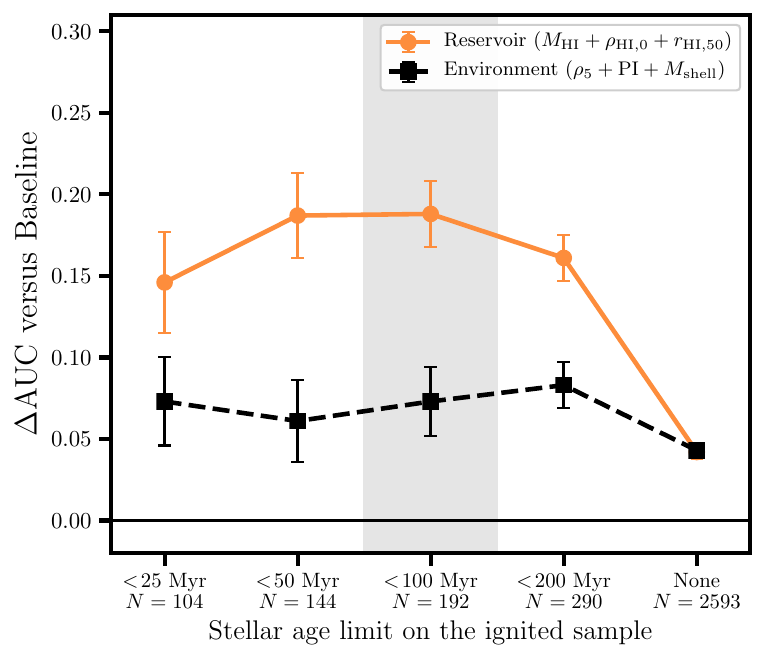}
\caption{Gain in AUC over the $\Mvir$, $\Mgas$, $z$ baseline when we vary the stellar age limit that defines the ignited sample. The starless class is the same $18{,}284$ halos in every column and $N$ is the number of ignited halos. The shaded column marks our sample ($<100$~Myr). The combined reservoir ($\MHI+\rhoHIo+r_{\rm HI,50}/\Rvir$) is shown in orange and the combined environment ($\rhofive+{\rm PI}+M_{\rm shell}$) in black. Error bars show the split-to-split scatter of each model's AUC over $100$ stratified splits.
\label{fig:agewindow}}
\end{figure}

Figure~\ref{fig:agewindow} shows that the combined HI reservoir adds more than the combined environment at every limit up to $200$~Myr. The margin is widest at $50$ and $100$~Myr, and with no limit the two families add the same amount. A paired bootstrap of the test set (\S\ref{subsec:regression}) gives a $95\%$ interval on the margin that excludes zero at $50$, $100$ and $200$~Myr but not at $25$~Myr, where only $104$ ignited halos remain. The ordering of the single properties is not fixed: $\MHI$ leads up to $100$~Myr ($+0.101$ against $+0.097$ at $100$~Myr), but at $200$~Myr $r_{\rm HI,50}/\Rvir$ exceeds it ($+0.104$ against $+0.071$). The reservoir's advantage therefore belongs to recently ignited halos, which is expected if the neutral-gas quantities describe the halo near the onset of star formation only while that event is recent. We therefore restrict the claims of this paper to recent ignition.

The same trend holds for halos hosting more than one star particle. Pooling halos with $N_\star=2$--$10$ star particles in total, all of them within $0.15\,\Rvir$ as for $N_\star=1$, and defining the ignited class by the age of the youngest of them, the reservoir's advantage over the environment grows as the limit tightens from $100$ to $25$~Myr, from a margin of $+0.051$ to $+0.072$ in AUC (the $95\%$ interval on this margin, from resampling the test set, excludes zero at every limit).

\subsection{The gas-particle floor}
\label{subsec:ngaswindow}
To test how the HI quantities depend on resolution, we restrict both classes to halos with at least $256$, $384$, $512$ or $1024$ gas particles, instead of our fiducial $128$, and repeat the regression. The floor also removes low-mass halos from both classes, so it tests resolution and halo mass together. It leaves $78$ of the $192$ ignited halos at $N_{\rm gas}=512$ and $18$ at $N_{\rm gas}=1024$.

\begin{figure}[tbp]
\centering
\includegraphics[width=\columnwidth]{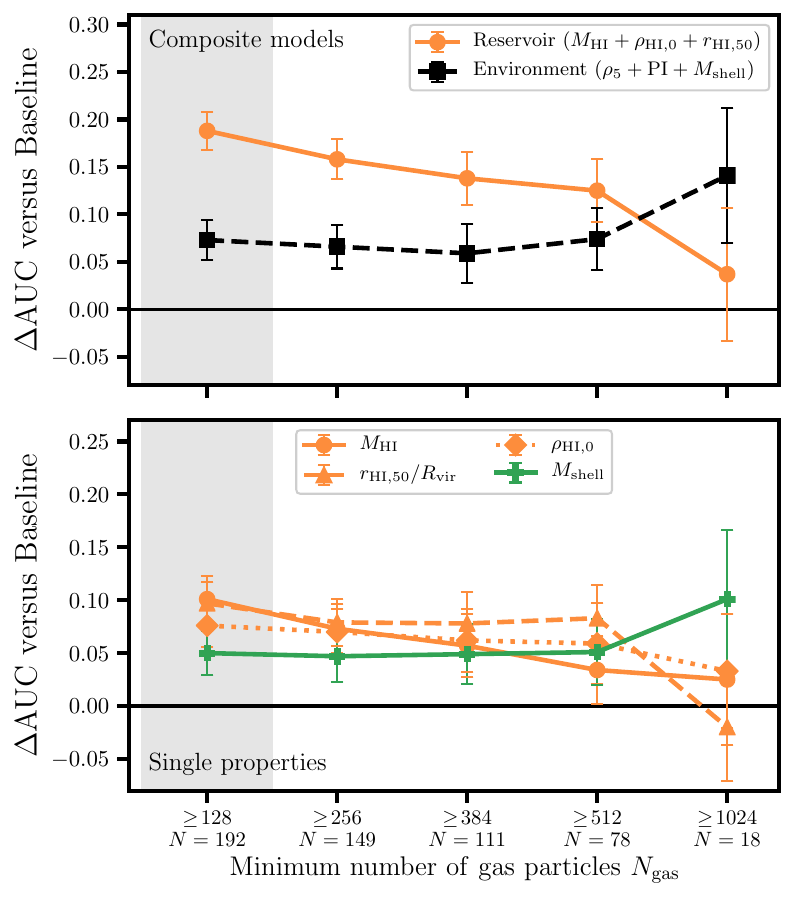}
\caption{Gain in AUC over the $\Mvir$, $\Mgas$, $z$ baseline when we restrict both classes to halos with at least $N_{\rm gas}$ gas particles. $N$ is the number of ignited halos and the starless class shrinks from $18{,}284$ to $520$ halos across the five columns. The shaded column marks our sample ($N_{\rm gas}\geq128$). \emph{Top:} the combined HI reservoir is shown in orange and the combined environment in black, as in Figure~\ref{fig:agewindow}. \emph{Bottom:} the three HI quantities one at a time (orange) and the shell mass $M_{\rm shell}$, the best single environmental property (green). The last column holds only $18$ ignited halos. Error bars are as in Figure~\ref{fig:agewindow}.
\label{fig:ngaswindow}}
\end{figure}

Figure~\ref{fig:ngaswindow} shows that the combined HI reservoir adds more than the combined environment up to $N_{\rm gas}=512$, although the margin narrows as the sample shrinks. The bootstrap interval on the margin excludes zero up to $N_{\rm gas}=256$ but not at higher gas-particle floors. Among the single properties (lower panel), $r_{\rm HI,50}/\Rvir$ keeps its gain up to $N_{\rm gas}=512$, $\rhoHIo$ declines gradually, and $\MHI$ alone falls below $M_{\rm shell}$ at $N_{\rm gas}=512$. At $N_{\rm gas}=1024$ the baseline AUC falls below $0.5$, so that column does not constrain the ranking.

The sign of the combined-reservoir advantage therefore persists as the floor rises, but its significance does not, because the sample becomes too small to resolve it above $N_{\rm gas}=256$. The standalone advantage of $\MHI$ is more sensitive to the floor, which may reflect the removal of low-mass halos rather than a numerical effect.

\begin{figure}[tbp]
\centering
\includegraphics[width=\columnwidth]{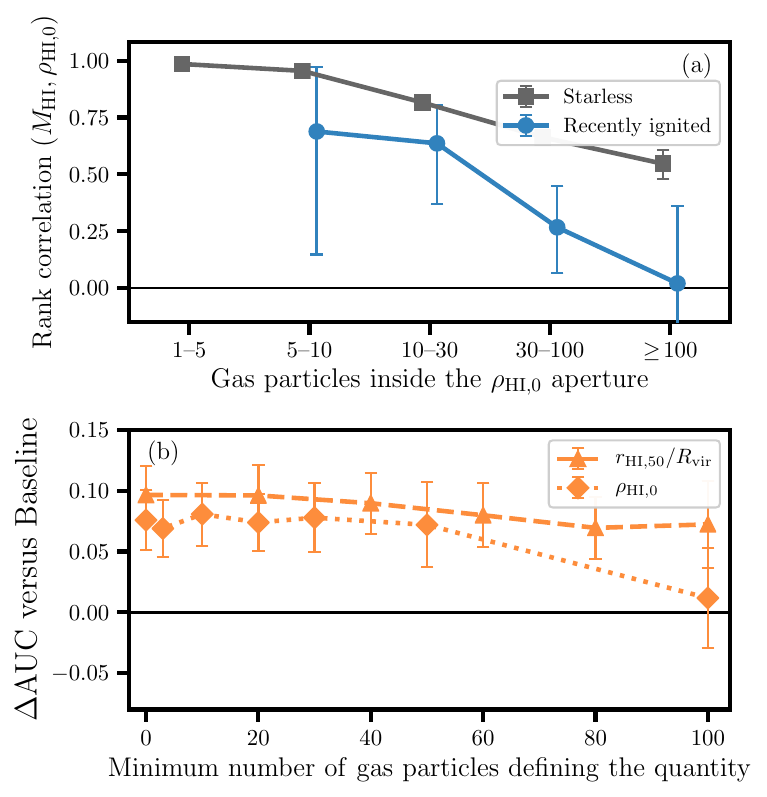}
\caption{Resolution of $\rhoHIo$ and $r_{\rm HI,50}/\Rvir$. \emph{Top:} rank correlation between $\MHI$ and $\rhoHIo$ versus the number of gas particles inside the $\rhoHIo$ aperture, for starless (gray) and recently ignited (blue) halos, with $95\%$ bootstrap intervals; bins with fewer than $15$ halos are not drawn. \emph{Bottom:} gain in AUC over the $\Mvir$, $\Mgas$, $z$ baseline of $r_{\rm HI,50}/\Rvir$ and $\rhoHIo$ when both classes are restricted to halos whose own measurement rests on at least the given number of gas particles (the particles interior to the half-mass radius for $r_{\rm HI,50}/\Rvir$, and those inside the aperture for $\rhoHIo$). Error bars are as in Figure~\ref{fig:agewindow}.
\label{fig:resolution}}
\end{figure}

The floor on $N_{\rm gas}$ does not necessarily guarantee that each HI quantity is itself resolved: halos have fewer gas particles inside the aperture that defines $\rhoHIo$. This bears on the decoupling of Section~\ref{subsec:intrinsic}. Among starless halos, the rank correlation between $\MHI$ and $\rhoHIo$ is dominated by halos with only a few particles in the aperture, and it falls steadily as the aperture fills (Figure~\ref{fig:resolution}a). Ignited halos, whose apertures hold more particles, fall faster, so the decoupling persists at matched particle number, although its size in the full samples is exaggerated by resolution.

We also restrict both classes to halos whose own measurement rests on at least a given number of gas particles (those interior to the HI half-mass radius for $r_{\rm HI,50}/\Rvir$, and those inside the $\rhoHIo$ aperture), and recompute the gain of each quantity over the baseline (Figure~\ref{fig:resolution}b). Both quantities keep most of their gain as the criterion tightens, except for $\rhoHIo$ in the last bin, where the ignited sample is too small to constrain it. The effective number of particles sharing the HI gives the same picture: the gain of $r_{\rm HI,50}/\Rvir$ is $+0.097$ without a criterion and $+0.069$ when at least $200$ effective particles share the HI ($100$ ignited halos remain). Because the criterion also removes low-mass halos, this also  tests resolution and mass together.

\subsection{The HI extent and the age of the first star particle}
\label{subsec:ageextent}
\begin{figure}[tbp]
\centering
\includegraphics[width=\columnwidth]{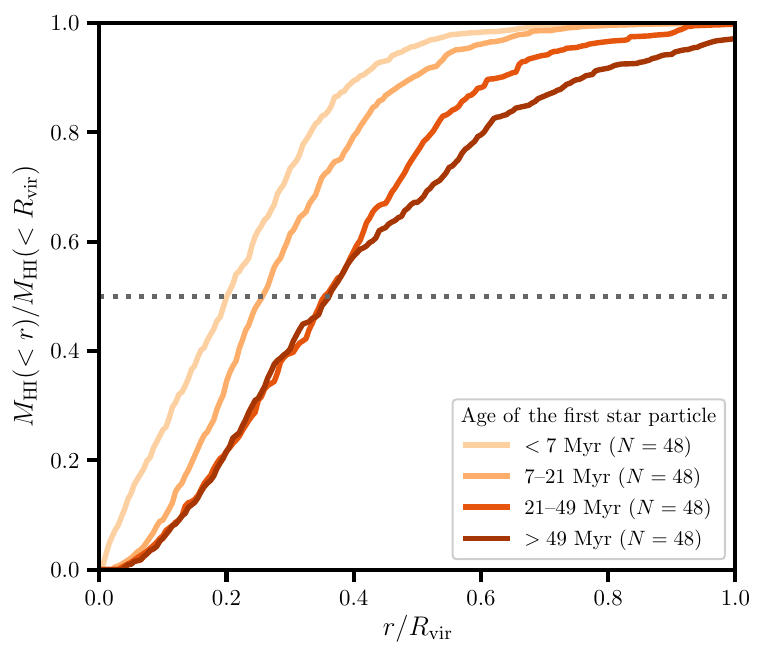}
\caption{Median cumulative HI mass profile of the ignited halos in four quartiles of the age of the first star particle ($N=48$ each), about the HI density peak and over the AHF-bound gas within $\Rvir$. The dotted line marks the half-mass radius.
\label{fig:ageextent}}
\end{figure}

Figure~\ref{fig:ageextent} shows the median cumulative HI profile of the ignited halos in four quartiles of the age of their first star particle. The profile shifts to larger radii with age: the median $r_{\rm HI,50}/\Rvir$ rises from $0.20$ in the youngest quartile to $0.36$ in the oldest, and the trend persists within $z=6$ alone. A natural question is whether this gas is expanding. The HI-weighted radial velocity inside $r_{\rm HI,50}$ is near zero in the youngest quartile and outward in the other three (medians of $+10$ to $+27\,{\rm km\,s^{-1}}$), as expected after the first stellar feedback, although its correlation with age is weak (Spearman $+0.13$ at fixed redshift). The radial extent is also only weakly correlated with the halo spin and concentration at fixed mass and redshift, and it adds to the regression beyond both, so it is not simply a proxy for either.

\clearpage
\bibliographystyle{aasjournalv7}
\bibliography{bibliography}

\end{document}